\documentclass[a4paper,11pt]{article}
\usepackage{jheppub} 

\def\CA{{\cal A}}
\def\CB{{\cal B}}
\def\CC{{\cal C}}
\def\CD{{\cal D}}

\def\CH{{\cal H}}

\def\CL{{\cal L}}

\def\CS{{\cal S}}
\def\CT{{\cal T}}

\usepackage{slashed}
\let\Slash\slashed

\newcommand{\gsim}{ \mathop{}_{\textstyle \sim}^{\textstyle >} }

\newcommand{\bra}[1]{ \langle {#1} | }
\newcommand{\ket}[1]{ | {#1} \rangle }
\newcommand{\et}[1]{  {#1} \rangle }

\def\diag{\mathop{\rm diag}\nolimits}
\def\Spin{\mathop{\rm Spin}}
\def\Pin{\mathop{\rm Pin}}
\def\pin{\mathop{\rm pin}}

\def\U{\mathrm{U}}

\def\tr{\mathop{\rm tr}}

\def\CA{{\cal A}}
\def\CC{{\cal C}}
\def\CD{{\cal D}}

\def\CF{{\cal F}}

\def\CH{{\cal H}}

\def\CL{{\cal L}}

\def\CS{{\cal S}}
\def\CT{{\cal T}}

\newcommand{\bC}{\mathbb{C}}
\newcommand{\bR}{\mathbb{R}}

\def\U{\mathrm{U}}

\def\O{\mathrm{O}}

\def\tr{\mathop{\mathrm{tr}}\nolimits}
\def\diag{\mathop{\mathrm{diag}}\nolimits}

\def\sign{\mathop{\mathrm{sign}}\nolimits}

\def\beq#1\eeq{\begin{align}#1\end{align}}

\usepackage{framed}
\definecolor{shadecolor}{rgb}{0.90,0.90,0.90}
\newenvironment{claim}{\begin{shaded}\noindent\itshape\ignorespaces}{\end{shaded}}

\usepackage{times}
\usepackage{courier}
\usepackage{mathtools}
\numberwithin{equation}{section}

\let\bar\overline
\def\ii{\mathrm{i}\,}

\def\dualPsi{{{\mathsf{\Psi}}}}

\def\dualX{{{\mathsf{X}}}}
\def\dualt{{{\mathsf{t}}}}

\title{
Dai-Freed theorem and topological phases of matter}

\preprint{IPMU-16-0094}

\author[]{Kazuya Yonekura}
\author[]{}
\affiliation[]{   Kavli IPMU (WPI), UTIAS, 
The University of Tokyo,  Kashiwa, Chiba 277-8583, Japan
}

\abstract{
We describe a physics derivation of theorems due to Dai and Freed about the Atiyah-Patodi-Singer eta-invariant
which is important for anomalies and topological phases of matter. 
This is done by studying 
a massive fermion.
The key role is played by the wave function of the ground state in the Hilbert space of the fermion in the large mass limit.
The ground state takes values in the determinant line bundle and has nontrivial Berry phases which characterize the low energy topological phases.
}

\begin{document} 
\maketitle
\flushbottom

\section{Introduction and summary}

What we call the Dai-Freed theorem~\cite{Dai:1994kq} is actually a set of theorems regarding Dirac operators on manifolds with boundary.
It has important implications for anomalies and topological phases of matter
\cite{Bruning:1996qx,Witten:1999eg,Harvey:1999as,Freed:1999vc,Moore:1999gb,
Bytsenko:2001ea,Grubb:2003yr,Diaconescu:2003bm,Asorey:2004kk,Freed:2004yc,Belov:2005ze,Freed:2006yc,Sati:2010mj,Park:2011cz,Dai:2011pk,Ibort:2012wi,
Monnier:2013rpa,Monnier:2014txa,Monnier:2014rua,Asorey:2015rxa,Witten:2015aba,Seiberg:2016rsg,Freed:2016rqq,Witten:2016cio,Bonora:2016nqc}.
In particular, the present paper is heavily influenced by \cite{Witten:2015aba,Witten:2016cio}.

\subsection{Description of the theorems}

\begin{figure}
\centering
\begin{tabular}{c@{\qquad}c}
\includegraphics[width=0.8\textwidth]{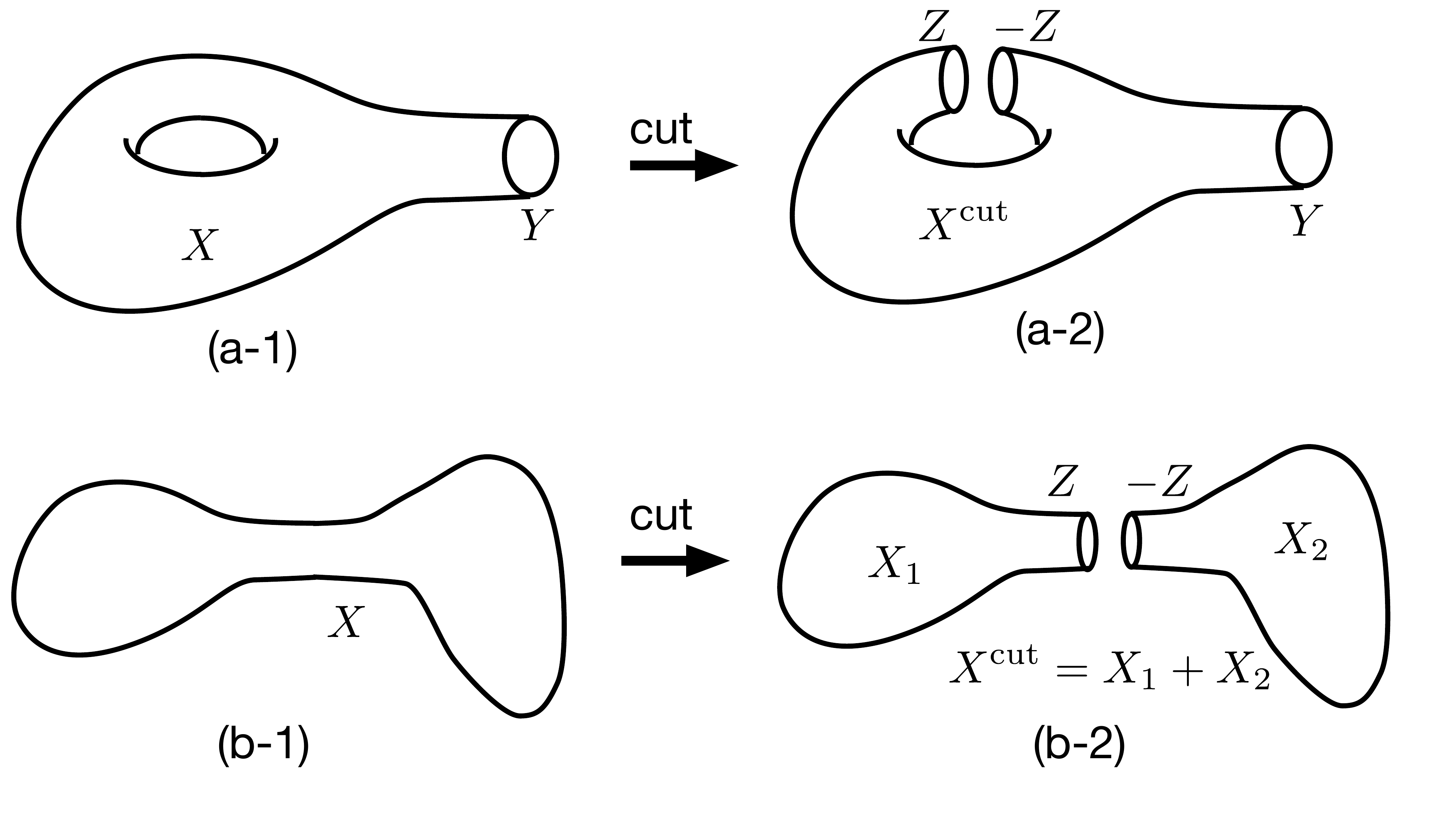}\\
\end{tabular}
\caption{Some examples of $X$ and $X^{\rm cut}$.
\label{fig:zu1}}
\end{figure}

Let $X$ be a $d+1$ dimensional manifold with boundary $\partial X =Y$. See the upper left of Fig.~\ref{fig:zu1} (a-1) for an example.
We remark that the boundary $Y$ (and also $X$ itself) 
is not necessarily connected and can have several connected components.
We assume that $X$ is equipped with a metric and some background gauge field 
(i.e., vector bundle $E$ with unitary connection)
which one specifies freely. In this paper we also assume that $X$ is either an odd dimensional spin manifold or an even dimensional $\pin^{\pm}$ manifold\footnote{
The $\Pin^{\pm}(n)$ groups are double cover of the orthogonal group $\O(n)$ whose connected component is $\Spin(n)$.
Then $\pin^\pm$ structures are uplifts of the structure group $\O(d+1)$ of the tangent bundle $TX$ to $\Pin^{\pm}(d+1)$ 
which are necessary to define fermions on unorientable manifolds.
See e.g., \cite{Kapustin:2014dxa,Witten:2015aba} for a review of $\Pin^{\pm}$ in the physics context.}
with the (s)pin bundle $S$.
In this setup, we can consider fermion fields (or more precisely sections $\Gamma( S \otimes E)$ of the bundle $S \otimes E$) coupled to the metric and background gauge field.
The Dirac operator is
\beq
\CD_X :=\ii \Slash{D}_X= \ii \gamma^\mu D_\mu
\eeq
where $\gamma^\mu$ are the gamma matrices and $D_\mu$ are the covariant derivative.

We want to consider the spectrum of the Dirac operator $\CD_X$, but this requires a careful preparation
for a manifold with boundary $\partial X=Y$ because of the problem of boundary conditions. 
The inner product between two fields $\Psi_1,\Psi_2 \in \Gamma( S \otimes E)$
in Euclidean signature is defined as
\beq
(\Psi_1, \Psi_2)_X=\int_X  \bar{\Psi}_1 \Psi_2,
\eeq
where the volume form $\sqrt{g}d^{d+1}x$ is implicit.
Near the boundary, we assume that the manifold is isometric to a cylinder
$(-\tau_0, 0] \times Y$, and the boundary is at $\tau=0$ where $\tau$ is the coordinate of $(-\tau_0, 0]$. 
Near the boundary, the Dirac operator is assumed to take the form
\beq
\CD_X= \ii \gamma^\tau \left( \frac{\partial}{\partial \tau}+\CD_Y \right) \label{eq:nearbd}
\eeq 
where $\CD_Y$ is a Dirac operator on the boundary $Y$.
Then by integration by parts, we get
\beq
(\Psi_1, \CD_X\Psi_2)_X - (\CD_X \Psi_1, \Psi_2)_X = \int_Y \bar{\Psi}_1 \ii \gamma^\tau \Psi_2  =   \ii ( {\Psi}_1,  \gamma^\tau \Psi_2 )_Y
\eeq
where $ ( {\Psi}_1,\Psi_2 )_Y :=  \int_Y \bar{\Psi}_1  \Psi_2 $.
This equation means that for the Dirac operator $\CD_X$ to be a hermitian operator, we have to impose a boundary condition 
such that the surface term $ ( {\Psi}_1,  \gamma^\tau \Psi_2 )_Y$ vanishes.
Furthermore, a boundary condition must be ``as weak as possible" while satisfying
this condition, because if we impose a too strong boundary condition (such as setting $\Psi|_Y=0$ at the boundary), there are no eigenmodes of $\CD_X$
that satisfy the boundary condition.\footnote{For example, one can check that the Dirac operator $\ii \gamma^\tau \frac{d}{d \tau}$
on the one-dimensional interval $X= [0,1]$ does not have any eigenmodes if we impose $\Psi=0$ at the boundary $\tau=0,1$.}
If $X$ is an odd dimensional spin manifold or an even dimensional $\pin^\pm$ manifold and if the fermion is in the irreducible
representation of the Spin/Pin group, one may convince oneself that there are no local boundary conditions consistent with the Lorentz symmetry. 
Therefore we must impose global boundary conditions
which we now describe.

The gamma matrix $\gamma^\tau$ in \eqref{eq:nearbd} satisfies $\gamma^\tau \CD_Y+\CD_Y \gamma^\tau=0$.
Thus $\gamma^\tau$ can be regarded as a chirality operator of the boundary Dirac operator 
$\CD_Y$. Therefore, on the boundary, we can split the fields into the positive and negative chirality parts as
\beq
\Psi|_Y = \Psi_++\Psi_-
\eeq
such that $\gamma^\tau \Psi_\pm= \pm \Psi_\pm$.
Correspondingly, the spin/$\pin^\pm$ bundle $S$ splits as $S|_Y = S_+ + S_-$ on the boundary, and
there are spaces of sections of the spin/pin bundles of positive and negative chirality coupled to the vector bundle
which we denote as $H_+(Y)=\Gamma(S_+ \otimes E|_Y)$ and $H_-(Y)=\Gamma(S_- \otimes E|_Y)$, respectively. These $H_\pm(Y)$ are infinite dimensional functional spaces.
The boundary term is now written as 
\beq
( {\Psi}_1,  \gamma^\tau \Psi_2 )_Y =( {\Psi}_{1,+},  \Psi_{2,+} )_Y - ( {\Psi}_{1,-},  \Psi_{2,- })_Y.
\eeq
This suggests the following boundary conditions. 
We pick up a unitary linear map 
\beq
T:  H_+(Y) \to H_-(Y),
\eeq
and impose the boundary condition given by
\beq
\Psi_- = T \Psi_+. \label{eq:bcd}
\eeq
Then the boundary term vanishes because $T$ is unitary:
$( T{\Psi}_{1,+},  T\Psi_{2,+} )_Y =( {\Psi}_{1,+},  \Psi_{2,+} )_Y$.
This boundary condition sets to zero only half of the $\Psi|_Y$ on the boundary. Setting at least half of the $\Psi|_Y$ to zero is required by the vanishing
of the boundary term. Thus it satisfies the condition of ``as weak as possible", and more precisely, $\CD_X$ is self-adjoint with this boundary condition
and it has well-defined spectrum (at least if the $T$ satisfies the condition below).

Because the behaviors of $\Psi_+$ and $\Psi_-$ under Lorentz transformations on $Y$ are different, the $T$ cannot be local in general.
One choice of $T$ is as follows. The boundary Dirac operator $\CD_Y$ splits into two parts based on chirality as
\beq
\CD_Y= \left( \begin{array}{cc}
0  & \CD_Y^{+-} \\
\CD_Y^{-+} & 0
\end{array}\right),
~~~~~\CD_Y^{-+}:  H_+(Y) \to H_-(Y),~~~~\CD_Y^{+-}:  H_-(Y) \to H_+(Y).
\eeq
If $\CD_Y$ does not have any zero modes, we can impose a boundary condition with $T=U_Y$, where we define
a unitary map as
\beq
U_Y = \frac{1}{\sqrt{\CD^{-+}_Y\CD^{+-}_Y}}\CD_Y^{-+}: H_+(Y) \to H_-(Y).\label{eq:sAPS}
\eeq
Generically, $\CD_Y$ does not have a zero mode because the $Y$ is the boundary of $X$ and in that case the index ${\rm Ind}(\CD_Y)$
is zero.\footnote{By Atiyah-Singer index theorem, the index ${\rm Ind}(\CD_Y)$ is given by the integral of a certain polynomial of curvatures which we denote as
$I_d$. Then ${\rm Ind}(\CD_Y) = \int_Y I_d =\int_X dI_d =0 $ because $I_d$ is closed.
 }
However, in the space of all possible metrics and gauge fields, 
there are points at which $\CD_Y$ gets zero modes, with the same number of the positive and negative chirality modes.
 These points are often guaranteed to exist by the arguments as in \cite{AlvarezGaume:1983cs,Witten:1982fp}.
 Thus we are led to consider more general boundary conditions.
We consider $T$ of the form
\beq
T= \left( \begin{array}{cc}
{T}|_{ \lambda_Y < \Lambda} & 0 \\
0 & U_Y|_{ \lambda_Y  \geq \Lambda}
\end{array}\right),
 \label{eq:gAPS}
\eeq
for some arbitrary $\Lambda>0$, 
where $|_{\lambda_Y<\Lambda}$ means that we are restricting to the subspace of $ H_\pm(Y) $ spanned by eigenmodes of $\CD_Y^2$ with eigenvalues $\lambda_Y^2<\Lambda^2$,
and the meaning of $|_{\lambda_Y \geq \Lambda}$ is similar.
The spaces $H_+(Y)|_{ \lambda_Y < \Lambda} $ and $H_-(Y)|_{ \lambda_Y < \Lambda} $ are finite dimensional, and 
 ${T}|_{ \lambda_Y < \Lambda}$ is an arbitrary unitary map between these spaces.
 The above condition means that the unitary map $T$ is basically arbitrary, except that for very high frequency modes $\lambda_Y \gg 1$
 the $T$ coincides with $U_Y$.
 
We call the boundary condition specified by $U_Y$ in \eqref{eq:sAPS} as the standard Atiyah-Patodi-Singer (APS) boundary condition~\cite{Atiyah:1975jf},\footnote{
In the case of the original APS setup, they had a chirality operator $\bar{\gamma}$ in the bulk $X$. In that case,
the standard APS boundary condition can be defined~\cite{Atiyah:1975jf} even if $\CD_Y$ has zero modes. 
In our case, we are not assuming the existence of $\bar{\gamma}$ in the bulk $X$.
Throughout the paper, the chirality means the one on the boundary $Y$ defined in terms of $\gamma^\tau$.}
and those specified by $T$ in \eqref{eq:gAPS} as generalized APS boundary conditions.

We have specified boundary conditions so that the $\CD_X$ is self-adjoint and has well-defined spectrum.
Denoting the eigenvalues of $\CD_X$ as $\lambda_X$, we define the APS eta-invariant $\eta_X(T)$ as
\beq
\eta_X(T):=\frac{1}{2} \left(  \sum_{\lambda_X \neq 0} {\rm sign}(\lambda_X)  + {\rm dim} {\rm Ker} \CD_X \right)_{\rm reg},
\eeq
where the sum is taken over all nonzero eigenvalues of $\CD_X$ including multiplicities, and the subscript ${\rm reg}$ means
some appropriate regularization which is usually done by zeta function regularization 
$ \eta_X(T,s) =  \frac{1}{2}(\sum_{\lambda_X \neq 0} {\rm sign}(\lambda_X)/|\lambda_X|^{s}  + {\rm dim} {\rm Ker} \CD_X )$.
This eta-invariant depends on the boundary condition $T$ and we made that dependence explicit in the notation $\eta_X(T)$.
We often abbreviate $\eta_X(T)$ just as $\eta(T)$ if the manifold $X$ is clear from the context.

Now we can state the first theorem.
The theorem is about the exponentiated eta-invariant $\exp(-2\pi \ii \eta(T))$.
Although $\eta(T)$ jumps discontinuously by integers when some eigenvalue $\lambda_X$ crosses zero, the $\exp(-2\pi \ii \eta(T))$
behaves smoothly under the change of metric and gauge field and so this is a natural quantity to consider.
Then we have\footnote{There is a few sign differences
between the formulas in this paper and those in \cite{Dai:1994kq}. This is due to a slight difference in the conventions.
In particular, our convention of APS boundary conditions is different from \cite{Dai:1994kq}.
Physically there is a natural convention for the standard APS boundary condition as we will explain in Sec.~\ref{sec:boundary}.}

\paragraph{Theorem 1.}
Let $T_1$ and $T_2$ be two unitary maps of the form \eqref{eq:gAPS} used in the boundary condition \eqref{eq:bcd}.
Then the exponentiated eta-invariant behaves as
\beq
\exp(-2\pi \ii \eta(T_2))=\det(T_2 T_1^{-1})\exp(-2\pi \ii \eta(T_1)) \label{eq:DFth1}.
\eeq
Here, $T_2 T_1^{-1}$ is of the form ${T}_2 {T}_1^{-1}|_{ \lambda < \Lambda} \oplus 1|_{ \lambda  \geq \Lambda}$
for some $\Lambda>0$ and the determinant is taken over the finite dimensional matrix ${T}_2 {T}_1^{-1}|_{ \lambda < \Lambda}$. \\

We need more preparation to state other theorems.
Suppose that $X$ has a codimension one submanifold $Z$ whose neighborhood in $X$ is given by a cylindrical region $(-\tau_0, \tau_0) \times Z$ for some $\tau_0$.  Then
let $X^{\rm cut}$ be the manifold which is obtained by
cutting $X$ along Z. If $\partial X = Y$, we have $\partial X^{\rm cut} = Y \sqcup Z \sqcup -Z$.
Here the minus sign on $-Z$ means that the definition of chirality on $-Z$ is opposite to that of $Z$, because of the change of coordinate $\tau \to -\tau$.
More explicitly, the chirality operator on $-Z$ is $\gamma^{-\tau}=-\gamma^\tau$ and the $\CD_X$ is 
$\CD_X = \gamma^{\tau} (\partial_\tau +\CD_Z) =  \gamma^{-\tau} (\partial_{-\tau} -\CD_Z)$, so we get $\CD_{-Z}^{-+} = - \CD_{-Z}^{+-}$.
In this case there are natural isomorphisms 
$H_\pm (-Z)  \cong H_\mp (Z)$.
We remark that we are assuming nothing about whether $X^{\rm cut}$ is connected or disconnected.
For example, we can consider a situation $X=S^1 \times Z$ and $X^{\rm cut}=[0,1] \times Z$.
Another example is the case that $\partial X=0$ and $X^{\rm cut} $ is a disconnected sum $X_1 + X_2$ where $\partial X_1 = - \partial X_2=Z$.
See Fig.~\ref{fig:zu1} for examples.

We impose a generalized APS boundary condition on $X^{\rm cut} $ such that 
the modes on $Y$ do not mix with the modes on $Z \sqcup -Z$ and hence it has the form $T_Y \oplus T_{Z \sqcup -Z}$.
More explicitly, we impose
\beq
(\Psi_- - T_Y \Psi_+)|_Y = 0,~~~~~(\Psi_- - T_{Z \sqcup -Z} \Psi_+)|_{Z \sqcup -Z} =0.
\eeq
A technical remark here is the following. The $Y$ is guaranteed to have vanishing index of $\CD_Y$ because it is the boundary of $X$ as mentioned before.
However, $Z$ is arbitrary and may have nonzero index of $\CD_Z$. In such a case, 
the dimensions of $H_+(Z)|_{ \lambda < \Lambda} $ and $H_-(Z)|_{ \lambda < \Lambda} $ are different and we cannot define a $T_Z$ which is unitary.
However, boundary conditions $T_{Z \sqcup -Z}$ on $Z \sqcup -Z$ are well-defined. If the index of $\CD_Z$ is zero,
it is also possible to impose generalized APS boundary conditions separately on $Z$ and $-Z$.

The second theorem is as follows.

\paragraph{Theorem 2.}
Under the cutting procedure $X \to X^{\rm cut}$ along $Z$, the exponentiated eta-invariant behaves as
\beq
\exp(-2\pi \ii \eta_{X^{\rm cut}}(T_Y\oplus T_{Z \sqcup -Z}))=  \det(T_{Z \sqcup -Z}) \exp(-2\pi \ii \eta_X(T_Y))  \label{eq:DFth2}
\eeq
Here, the determinant $\det(T_{Z \sqcup -Z})$ is taken by using the natural isomorphism
$H_+(Z \sqcup -Z) \cong H_+(Z) \oplus H_+(-Z) \cong H_-(-Z) \oplus H_-(Z) \cong H_-(Z \sqcup -Z) $.

\paragraph{Corollary.}
If $\partial X=\varnothing$ and $X^{\rm cut}=X_1 + X_2$ with $\partial X_1 = -\partial X_2=Z$, 
we have
\beq
\exp(-2\pi \ii \eta_{X_1}(T_Z)  ) \exp(-2\pi \ii \eta_{X_2}(-T_{Z}^\dagger)) = \exp(-2\pi \ii \eta_X)   \label{eq:DFth2-1}
\eeq 
Here $T_{Z}^\dagger:H_-(Z) \to H_+(Z)$ is regarded as $T_{Z}^\dagger:H_+(-Z) \to H_-(-Z)$. 
Notice that the standard APS boundary condition satisfies $U_{-Z}= - U_Z^\dagger$ because $\CD_{-Z}^{-+} = -\CD_Z^{+-}$.\\

Let us proceed to the third theorem. In \eqref{eq:DFth1} we have seen the $T$ dependence of the exponentiated eta-invariant $\exp(-2\pi \ii \eta_X(T)) $.
This suggests us to consider the following quantity,
\beq
\CT(X) := \frac{\exp(-2\pi \ii \eta_X(T)) }{ \det T}.\label{eq:tausection}
\eeq
By Theorem~1, this is independent of $T$. However, this quantity is not naturally a numerical number, but takes values in a one dimensional vector space.
Remember that $T$ is a map $H_+(Y) \to H_-(Y)$ where $Y =\partial X$. Then the inverse of the determinant $\det T$ and hence $\CT(X)$
naturally takes values in a one-dimensional vector space 
\beq
 \det H_+(Y) \otimes \det H_-(Y)^*, 
\eeq
where, in general for a given vector space $V$, the notation $\det V$ means the one-dimensional vector space given by the top exterior product
$\bigwedge^{\dim V} V$, and $V^*$ is the dual space of $V$. The spaces $H_+(Y) $ and $ H_-(Y)$ are infinite dimensional, but for high frequency modes $\lambda_Y  \geq \Lambda$,
 there is the isomorphism given by $U_Y|_{\lambda_Y \geq \Lambda}$, and hence these infinite dimensional spaces 
 are effectively reduced to be finite dimensional in the above determinant. In other words, we can just consider 
 $ \det H_+(Y) \otimes \det H_-(Y)^* \cong \det H_+(Y)|_{\lambda_Y <\Lambda} \otimes \det H_-(Y)^*|_{\lambda_Y <\Lambda}$ for some arbitrary $\Lambda>0$.
 
 If $U_Y$ were always well-defined even for modes with small eigenvalues, we could have trivialized $ \det H_+(Y) \otimes \det H_-(Y)^*$ completely by using $U_Y$. 
 However, $U_Y$ becomes ill-defined when some eigenvalues of $\CD_Y$ become zero.
 More precisely, let us consider a fiber bundle $F$ over some base $W$. We call the base a parameter space.
 The typical fiber of $F$ is $X$, and the metric and gauge field on $X$ vary as we move the parameter space $W$, meaning that $W$ parametrizes metric and gauge field on $X$.
 We denote the situation as
 \beq
  \pi: F \to W,~~~ 
  \pi^{-1}(w)=X_w,~~\partial X_w = Y_w~~(w \in W).
 \eeq
 The metric and gauge field are assumed to be extended to the total space $F$ in an appropriate way.\footnote{
 More precisely, the metric on the total space of the bundle $F$ is assumed to be of the form 
\beq
 ds^2=g_{\mu\nu}(x,w) (dx^\mu - B^\mu_a(x,w) d w^a)(dx^\nu - B^\nu_b(x,w) d w^b)+\frac{1}{\epsilon^2}g_{ab}(w)dw^a dw^b,
 \eeq 
 where $x^\mu$ and $w^a$ are coordinates of the fiber $X$ and the base $W$, respectively.
 We take $\epsilon \to 0$ at the end. This limit is called the adiabatic limit. 
 The horizontal distribution of the fiber bundle $F$, defined by $dx^\mu - B^\mu_a(x,w) d w^a$, is a part of the data of Theorem~3,
 but we suppress this dependence in this paper. See \cite{Freed:1986hv} for more details.}
Then we can define a line bundle
\beq
\CL \to W ,~~~~~~~~
\CL_w = \det H_+(Y_w) \otimes \det H_-(Y_w)^*.
\eeq
This line bundle is called the determinant line bundle of $\CD_Y$ (see e.g., \cite{Freed:1986hv}).
Notice that this line bundle only depends on $Y_w$, and not on $X_w$. From the above consideration, we see that
the obstruction for trivializing $\CL$ comes from zero eigenvalues of the boundary Dirac operator $\CD_Y$.

The $\CT(X_w) $ takes values in $\CL_w$,
\beq
\CT(w):=\CT(X_w) \in \CL_w
\eeq
and hence $\CT$ defines a section of the line bundle $\CL$.
Then we have

\paragraph{Theorem 3.} There exists a natural connection $\nabla$ of the determinant line bundle $\CL$.
Under this connection, $\CT$ behaves as
\beq
\nabla \CT(w) = \CT(w) \cdot \int_{X_w} 2 \pi \ii I_{d+2}.  \label{eq:DFth3}
\eeq
Here, $I_{d+2}$ is a $d+2$-form given by
\beq
I_{d+2}= \left. \hat{A}(R) \tr \exp( \frac{\ii F}{2\pi}) \right|_{d+2},
\eeq
where $\hat{A}(R)$ is the $\hat{A}$-genus of the metric and $F$ is the curvature tensor of the gauge field.
Notice that $\int_{X_w} I_{d+2}$ is the integral of a $d+2$-form on $d+1$-dimensional manifold $X_w$ and hence it gives a 1-form on the base $W$.
For odd $d$ with $\pin^\pm$ structures (i.e., unorientable), we define $I_{d+2}=0$.

\subsection{Summary of the paper}
It have taken us a long preparation for just stating the theorems, but it is worth it.
The importance for anomalies is reviewed in Appendix~\ref{sec:appA}.
In the rest of this paper, we will show that the ingredients of the Dai-Freed theorem naturally appear in the study of a massive fermion in $d+1$ dimensions
\beq
\CL = - \bar{\Psi} (-\ii \CD_X+m) \Psi
\eeq
and its low energy topological phases.
The difference of our work from \cite{Witten:2015aba,Witten:2016cio}
is that we study the case where the boundary $Y=\partial X$ is regarded as a time slice, whereas 
in \cite{Witten:2015aba,Witten:2016cio} the case where $Y$ is a spatial boundary has been considered.

If the low energy topological phase of the theory with $m>0$ is trivial,
the theory with $m<0$ is nontrivial. Then, for $m<0$, the results of this paper may be summarized as follows.
\begin{claim}
\begin{itemize}
\item { Sec.~\ref{sec:boundary}:} Generalized APS boundary condition with respect to $T$ give a physical state $\ket{T}$ in the Hilbert space 
$\CH_Y$ of the massive fermion on $Y$. 
For the ground state $\ket{\Omega}$, we can compute the ``wave function of the ground state" $\bra{T}\et{\Omega}$ as a function of $T$. 
This is somewhat analogous to wave functions $\bra{x}\et{\Omega}$ in the usual quantum mechanics
in the coordinate basis $\ket{x}$. The wave function is given by $\bra{T}\et{\Omega} \sim \det{T}$ in the large mass limit.

\item { Sec.~\ref{sec:DF}:} The exponentiated eta-invariant is given by the amplitude $\exp(-2\pi \ii \eta(T)) \sim \bra{T}\et{X}$, where $\ket{X} \in \CH_Y$ is the state
obtained by the path integral on $X$. The theorems follow from the fact that the Euclidean path integral is dominated by the ground state $\ket{\Omega}$
in the large mass limit. 

\item { Sec.~\ref{sec:Berry}:} The ground state 
takes values in the determinant line bundle over the parameter space of metric and gauge field.
There are natural parallel transport, connection, and curvature in the determinant line bundle defined by Euclidean path integral.
They give Berry phase, connection and curvature of the ground state. The $\CT(X)$ is naturally identified with $\ket{X}$.
\end{itemize}
\end{claim}

\section{Physical states and boundary conditions of the path integral}\label{sec:boundary}
Let us quantize a massive fermion on a cylinder $X= \bR \times Y$, where $\bR$ is the time direction.
The coordinate of the time direction is denoted as $t$ or $\tau$ in Lorenzian or Euclidian signature, respectively,
where $t= -\ii \tau$.
We assume that the time direction is completely flat.

We consider a Dirac fermion for simplicity, but when the $\Spin$/$\Pin^\pm$ and the gauge group permit a majorana fermion, further refinement
is possible which is essentially the square roots of the formulas. However, we do not perform the analysis explicitly for the majorana case.

Let $\CD_X=\ii \Slash{D}_X$ and $\CD_Y$ be the Dirac operators on $X$ and $Y$ as in the Introduction.
The Lagrangian of the fermion $\Psi$ is
\beq
\CL = - \bar{\Psi} (-\ii \CD_X+m) \Psi=\ii \Psi^\dagger \partial_t \Psi -\Psi^\dagger ( \CD_Y+m \gamma^\tau) \Psi,
\eeq
where $\bar{\Psi}=\Psi^\dagger \gamma^\tau$. The Hamiltonian $H_Y$ is
\beq
H_Y=\int_Y \Psi^\dagger ( \CD_Y+m \gamma^\tau) \Psi.
\eeq
Assuming that the index of $\CD_Y$ is zero, the eigenmodes of $\CD_Y$ form pairs.
We denote a pair labeled by $i$ as $(\psi_{+,i},\psi_{-,i})$ which satisfy
\beq
\gamma^\tau \psi_{\pm,i}=\pm \psi_{\pm,i},~~~\CD_Y \psi_{\pm,i}=\lambda_{Y,i}  \psi_{\mp,i}~~(\lambda_{Y,i} \geq 0).
\eeq
Note that $ \psi_{+,i} \pm  \psi_{-,i}$ are eigenmodes of $\CD_Y$ with eigenvalues $\pm \lambda_{Y,i}$, respectively.
For zero modes there is no natural choice for the parings of $\psi_{+,i}$ and $\psi_{-,i}$, but anyway we just choose some pairings for convenience of the following analysis.
Then, the fermion can be expanded as
\beq
\Psi = \sum_i \left(   A_{+,i} \psi_{+,i} + A_{-,i} \psi_{-,i}  \right) \label{eq:modeexp}
\eeq
and the Hamiltonian is 
\beq
H_Y=\sum_i (A_{+,i}^\dagger, A_{-,i}^\dagger) \left( \begin{array}{cc}
m & \lambda_{Y,i} \\
\lambda_{Y,i} & -m
\end{array} \right)
 \left( \begin{array}{c}
A_{+,i}  \\
A_{-,i}
\end{array} \right).
\eeq
The coefficients satisfy the canonical anti-commutation relations
\beq
\{ A_{+,i}^\dagger, A_{+,j} \}=\{ A_{-,i}^\dagger, A_{-,j} \}=\delta_{ij},
\eeq
and other anti-commutators are zero.

\subsection{The relation of states and boundary conditions}\label{sec:QM}
Before considering the above fermion system, let us first consider a simple quantum mechanical fermion system
on $1+0$ dimension $\bR$ with the Lagrangian $L=\ii Q^\dagger \partial_t Q - H$, where $H$ is the Hamiltonian. The anti-commutation relation is
$\{Q^\dagger, Q \}=1$ and there are two states $\ket{ \pm}$ characterized by $Q \ket{-}=0$ and $\ket{+}=Q^\dagger \ket{-}$.
We regard $Q$ as the canonical coordinate and $P=\ii Q^\dagger$ as the canonical momentum $P=\ii \frac{\partial}{ \partial Q}$.

The fermion path integral is based on coordinate and momentum eigenstates~\cite{Weinberg:1995mt,Polchinski:1998rq}. We consider states satisfying
\beq
&Q \ket{q} = \ket{q}q  ,~~~~~~\bra{q} Q=  -q\bra{q} , \\
&P\ket{p}= -\ket{p} p, ~~~~\bra{p}P =p\bra{p} ,
\eeq
where $q,p$ are grassmann variables.
These can be realized by taking
\beq
&\ket{q}=\ket{-} + \ket{+}q  ,~~~~~~~~\bra{q}= q\bra{-} -\bra{+} , \\
&\ket{p}= -\ket{-} \ii p - \ket{+}  ,~~~~\bra{p}= \bra{-}  - \ii p \bra{+} .
\eeq
The above definitions may look strange, but they are chosen to avoid the issue of whether $\ket{-}$ is bosonic or fermionic,
that is, whether $q \ket{-}=+\ket{-}q$ or $q \ket{-}=-\ket{-}q$. If $\ket{-}$ is bosonic, we simply have
$Q \ket{q} = q\ket{q},~ \bra{q} Q=  \bra{q}q, ~ P\ket{p}= p\ket{p} $ and $\bra{p}P =\bra{p}p$.

These states are chosen to satisfy the orthogonality relations
\beq
\bra{q}\et{q'}=q-q'=\delta(q-q'),~~~~~\bra{p}\et{p'}=\ii (p-p') =\ii \delta (p-p'),
\eeq
where $\delta(q)=q$ is the delta function for grassmann variables. They also satisfy the Fourier transform relations
\beq
&\bra{q}\et{p}=e^{ \ii p q},~~~\bra{p}\et{q}=e^{-\ii pq}.
\eeq
We have the following completeness relations
\beq
&\int \ket{q} dq \bra{q} = \int  \ket{p} (- \ii dp) \bra{p}= \ket{-} \bra{-} +\ket{+} \bra{+} ={\bf 1},
\eeq
where ${\bf 1}$ is the identity operator acting on the Hilbert space.

An infinitesimal evolution of time $d \tau$ in Euclidian signature is described as
\beq
 e^{-H d \tau}\ket{q} &=\int  \ket{q'} dq' \bra{q'} \int  \ket{p} (- \ii dp) \bra{p}  e^{-H d\tau}\ket{q} \nonumber \\
 &=\int \frac{dq' dp}{\ii}  \ket{q'} e^{\ii p d q - H(p,q) d\tau}
\eeq
where $d q=q'-q$, and $H(p,q)$ is the Hamiltonian evaluated between $\bra{p}$ and $\ket{q}$. 
By using these formulas, the path integral is derived in the standard way.

For our purposes, the important point is as follows. Let $\ket{\alpha}$ be an arbitrary state, and consider amplitudes $\bra{i} e^{-H \tau_o} \ket{\alpha}~(i=\pm)$.
The ``last step" in the path integral is given by
\beq
\bra{i} e^{-H \tau_o} \ket{\alpha}=  \int \bra{i}\et{q} dq  \bra{q}e^{-H \tau_o} \ket{\alpha}.\label{eq:finaltime}
\eeq
The inner product $\bra{i}\et{q}$ is given by
\beq
\bra{-}\et{q}=1,~~~~~~\bra{+}\et{q}=q=\delta(q).\label{eq:simpleboundary}
\eeq
These formulas \eqref{eq:finaltime} and \eqref{eq:simpleboundary} mean the following.
When we try to compute $\bra{i} e^{-H \tau_o} \ket{\alpha}$ by the path integral, the boundary conditions at the final time slice are such that the canonical coordinate
$q$ is unconstrained if $\bra{i}=\bra{-}$ and it is constrained to be zero $q=0$ if $\bra{i}=\bra{+}$ by the delta function $\delta(q)$.
In the same way, the canonical momentum $p$ is constrained to be zero $p=0$ for $\bra{-}$ and unconstrained for $\bra{+}$.

Conversely, if we compute the path integral with the boundary condition that $q=0$ with $p$ unconstrained at the final time,
that corresponds to computing an amplitude with the final state $\bra{+}$. In the same way, 
if we compute the path integral with the boundary condition that $p=0$ with $q$ unconstrained at the final time,
that corresponds to computing an amplitude with the final state $\bra{-}$. 
These statements can be generalized to multi-variable cases in the obvious way.

\subsection{Generalized APS boundary conditions}
Let us return to the fermion on $X =\bR \times Y$.
Generalized APS boundary conditions are defined as follows. We use the basis defined by the mode expansion \eqref{eq:modeexp}.
Let $H_+(Y)$ be the space spanned by $\psi_{+,i}$, and let $H_-(Y)$ be the space spanned by $\psi_{-,i}$.
Let $T=(T_{ij})$ be a unitary matrix from $H_+(Y)$ to $H_-(Y)$ 
such that $T_{ij}=\delta_{ij}$ for modes $\psi_{\pm, i}$ of large enough eigenvalues $\lambda_{Y,i} $. 
As long as this condition for high frequency modes is satisfied, the unitary matrix $T=(T_{ij})$ is really arbitrary.

We denote $\vec{A}_\pm = (A_{\pm,,i})$ and then define
\beq
\vec{B}_1=\frac{\vec{A}_- - T\vec{A}_+}{\sqrt{2}},~~~\vec{B}_2=\frac{\vec{A}_- + T\vec{A}_+}{\sqrt{2}}. \label{eq:APSmode}
\eeq
Then, the generalized APS boundary condition defined by $T$ is such that $\vec{B}_1=0$ and $\vec{B}_2$ is unconstrained.

For $\Psi^\dagger$, we need to be a little bit careful. As an operator acting on the Hilbert space $\CH_Y$ associated to the space $Y$,
the hermitian conjugate of $\Psi$ is $\Psi^\dagger$. However, for the problem of determining the spectrum of the Dirac operator $\CD_X$
on Euclidean spaces $X$, the $\bar{\Psi}=\Psi^\dagger \gamma^\tau$ is considered as
living on the space of sections conjugate to that of $\Psi$. Moreover, $\Psi$ and $\bar{\Psi}$ are treated as independent variables in the path integral.
For $\bar{\Psi}$, the generalized APS boundary condition is such that $\int_Y \bar{\Psi}\gamma^\tau \Psi$
is zero for any $\Psi$ satisfying the above boundary condition. In terms of $\Psi^\dagger$, this means that 
$\int_Y \Psi^\dagger \Psi$ is zero. One can see that this condition is equivalent to saying that when a canonical coordinate $Q$ is unconstrained,
then the corresponding canonical momentum $P$ is set to zero at the boundary and vice versa. This is exactly as we have seen in
Sec.~\ref{sec:QM}.

Computing the path integral with the generalized APS boundary condition $T$ corresponds to computing the amplitude with a certain final state which we denote as $\bra{T}$
(where $\bra{T} \in \CH_Y^*$).
From the above discussion of the single variable case, it is clear that $\bra{T}$ must satisfy the conditions
\beq
\bra{T}\vec{B}_1=0,~~~~~\bra{T}\vec{B}_2^\dagger=0
\eeq
or equivalently
\beq
\vec{B}_1^\dagger \ket{T}=0,~~~~~\vec{B}_2\ket{T}=0. \label{eq:gAPSstate}
\eeq

\paragraph{The standard APS boundary condition.}
The standard (as opposed to generalized) APS boundary condition is given by $T_{ij}=\delta_{ij}$, which corresponds to 
$T=U_Y$ in the basis independent notation \eqref{eq:sAPS}.
This boundary condition is naturally defined only when the Dirac operator $\CD_Y$ does not have any zero modes.

In this case, we have $\vec{B}_1=(\vec{A}_- - \vec{A}_+)\sqrt{2} $ and $\vec{B}_2=(\vec{A}_- + \vec{A}_+)\sqrt{2} $.
If the mass of the fermion is neglected, the standard APS boundary condition has a very natural interpretation.
The Hamiltonian is given by
\beq
H_Y= \sum_i \left( -\lambda_{Y,i} B_{1,i}^\dagger B_{1,i}+ \lambda_{Y,i} B_{2,i}^\dagger B_{2,i} \right) +\text{mass term}.
\eeq
Therefore, the APS condition \eqref{eq:gAPSstate} means that the state $\ket{T=U_Y}$ is the ground state of the massless theory $m=0$.

Even in generalized APS boundary conditions, we require that $T_{ij}=\delta_{ij}$ for large $\lambda_{Y,i}$. 
This is also physically natural. For very high frequency eigenmodes $\psi_{\pm,i}$ with large $\lambda_i $, 
we want the state $\ket{T}$ to be at the unexcited states of these high frequency modes to avoid infinitely large energy.
At least for the modes $\lambda_{Y,i} \gg |m|$, this is achieved by the condition $T_{ij}=\delta_{ij}$.

\subsection{The wave function of the ground state}
In general, for nonzero mass $m$ or for generic $T$, the state $\ket{T}$ is not the ground state. 
The ground state $\ket{\Omega}$ is given by the following conditions. First, define $\theta_{i, m}$ by
\beq
(\cos 2\theta_{i, m}, \sin 2\theta_{i, m})=\frac{(m, \lambda_{Y,i})}{\sqrt{m^2+\lambda_{Y,i}^2}}~~~~~(0 \leq \theta_{i, m} \leq \frac{\pi}{2}).
\eeq
Then we define
\beq
{C}_{1,i} = -\sin \theta_{i, m} {A}_{+,i} +\cos \theta_{i,m} {A}_{-,i} ,~~~~~{C}_2 =   \cos \theta_{i, m} {A}_{+,i} +\sin \theta_{i,m} {A}_{-,i},
\eeq
or by using a matrix notation, we write
\beq
\vec{C}_{1} = -\CS_m \vec{A}_{+} + \CC_m \vec{A}_{-} ,~~~~~\vec{C}_2 =   \CC_m \vec{A}_{+} + \CS_m \vec{A}_{-},
\eeq
where $\CC_m=\diag(\cos \theta_{i,m}) $ and $\CS_m=\diag(\sin \theta_{i,m})$. The ground state is given by
\beq
\vec{C}_1^\dagger \ket{\Omega_m}=0,~~~~~\vec{C}_2 \ket{\Omega_m}=0.
\eeq
The relation between $\vec{B}_{1,2}$ and $\vec{C}_{1,2}$ is given by 
\beq
\left( \begin{array}{c}
\vec{B}_1 \\
\vec{B}_2
\end{array} \right)
=\frac{ 1 }{ \sqrt{2} }
\left(\begin{array}{cc}
\CC_m+T\CS_m, & \CS_m - T \CC_m \\
\CC_m-T\CS_m, & \CS_m + T \CC_m
\end{array} \right)
\left( \begin{array}{c}
\vec{C}_1 \\
\vec{C}_2
\end{array} \right).\label{eq:BCtransition}
\eeq

The overlap $\bra{T}\et{\Omega_m}$ is 
formally computed as follows. We pretend as if the Hilbert space is finite dimensional.
Let $\ket{E}$ be the state satisfying $\vec{B}_{1}\ket{E}=\vec{B}_{2}\ket{E}=0$, which also imply 
$\vec{C}_{1}\ket{E}=\vec{C}_{2}\ket{E}=0$. Then the states $\ket{T}$ and $\ket{\Omega_m}$ are given as
$\ket{T}=\prod_i B_{1,i}^\dagger \ket{E}$ and  $\ket{\Omega_m}=\prod_i C_{1,i}^\dagger \ket{E}$. Thus
we get $\bra{T}\et{\Omega_m}=\bra{E}\prod^{\rm rev}_i B_{1,i} \prod_i C_{1,i}^\dagger \ket{E}$, 
where $\prod^{\rm rev}_i $ means that the order of the product is reversed from that of $\prod_i$. 
By substituting \eqref{eq:BCtransition},
we get the result
\beq
\bra{T}\et{\Omega_m} = \det \left[ \frac{\CC_m+T\CS_m}{\sqrt{2}} \right].
\eeq
This formal expression actually needs regularization because of the infinite product in the determinant, but
one can check that a kind of Pauli-Villars regularization is possible. Alternatively, the ratio $\bra{T}\et{\Omega_{m=-m_0}}  /  \bra{T}\et{\Omega_{m=m_0}}$ 
between the theories with negative and positive mass 
is well-defined, and that is enough for our purposes.
In any case, we assume that some regularization is done and we neglect very high frequency modes.

This product $\bra{T}\et{\Omega_m}$ may be regarded as the ``wave function of the ground state".
In the usual quantum mechanics, wave functions such as $\bra{x}\et{\Omega}$ are computed by imposing the boundary condition 
that $x(t)|_{t=t_f}=x$ at the final time $t_f$. In our case, the $\bra{T}$ was obtained by generalized APS boundary conditions.
In this respect, $\bra{T}\et{\Omega_m}$ as a function of $T$ can be considered as the wave function.
However, we remark that $\bra{T}$ are not linearly independent if we consider all possible $T$.

Now let us consider the limit $m \to \pm \infty$. More precisely, we assume that the 
modes with eigenvalues $\lambda_i$ comparable to or larger than $|m|$, that is $\lambda_i \gsim |m|$, have the standard 
boundary condition $T_{ij}=\delta_{ij}$. Then, for the purpose of considering the dependence of $\bra{T}\et{\Omega_m}$ on the nontrivial part of $T$,
we can consider $\lambda_i/|m| \ll 1$ and neglect it.
Now we need to distinguish two cases. The first case is $m>0$. In this case, $\theta_{i,m} \to 0$ as $m \to +\infty$ and hence we get
\beq
\bra{T}\et{\Omega_m} \to (\text{const.})~~~(m \to +\infty)
\eeq
after a suitable regularization, where $(\text{const.})$ means that it is independent of $T$.

The second case is $m<0$ and this case is more interesting. We have $\theta_{i,m} \to \pi/2$ as $m \to -\infty$,
and hence we get
\beq
\bra{T}\et{\Omega_m} \to (\text{const.}) \cdot \det(T) ~~~(m \to -\infty).
\eeq
By using the fact that $\theta_{i, -m_0} = \pi/2 - \theta_{i, m_0}$, one can check that the overall constant appearing here is the same as the one appearing in the case $m>0$.
Therefore, we finally get the result
\beq
\lim_{m_0 \to \infty}\frac{\bra{T}\et{\Omega_{-m_0}}}{\bra{T}\et{\Omega_{m_0}}}=\det(T).\label{eq:ratiodetT}
\eeq
This is the crucial result for the $T$ dependence of the eta-invariant $\eta(T)$ as we will see.

We remark that the $\det(T)$ here is taken with respect to the explicit basis chosen in \eqref{eq:modeexp}.
Under a change of the basis, the $\det(T)$ changes, but that is not a problem because the phase factor of $\ket{\Omega_m}$ 
also depends on the choice of basis. We will give more systematic discussion in Sec.~\ref{sec:Berry}.

\section{The path integral on a manifold with boundary}\label{sec:DF}

Now we can give a proof of Theorem~1 and Theorem~2 stated in the Introduction. 
Let $X$ be a $d+1$ dimensional manifold with spin or $\pin^{\pm}$ structure and background gauge field.
The boundary of $X$ is denoted as $\partial X =Y$. Near the boundary, we assume that the manifold is isometric to a cylinder
$(-\tau_0, 0] \times Y$ where $\tau_0$ is a constant and the boundary is at $0 \in (-\tau_0, 0]$. The complement of this cylindrical part in $X$ is denoted as $X'$.
Thus $X= X' \cup [-\tau_0, 0] \times Y$ which are glued along $\partial X'$ and $\{ -\tau_0 \} \times Y$.
We do not assume anything about whether $Y$ is connected or not.

\subsection{Derivation of Theorem 1}
Regarding $Y=\partial X$ as a time-slice, the path integral of the fermion on the manifold $X$ gives a state in the Hilbert space $\CH_Y$ on $Y$ which we denote as 
$\ket{X} \in \CH_Y$. First we show (following \cite{Witten:2015aba}) that the amplitude $\bra{T} \et{X}$ is related to the eta-invariant $\eta(T)$ as
\beq
\lim_{m_0 \to \infty} \frac{\bra{T} \et{X}_{ m= - m_0}}{\bra{T} \et{X}_{ m= + m_0}} = e^{-2 \pi \ii \eta(T)}
\eeq
where the numerator and denominator are the amplitudes in the theories with the mass parameter given by $m=-m_0$ and $m=+m_0$, respectively.
This is shown as follows.
We showed in the previous section that the amplitude $\bra{T} \et{X}$ is given by the path integral with the generalized APS boundary condition specified by $T$.
Thus we get
\beq
\frac{\bra{T} \et{X}_{ m= - m_0}}{\bra{T} \et{X}_{ m= + m_0}}  = \frac{\det( -\ii \CD_X -m_0)}{\det(-\ii \CD_X +m_0)} = \prod_{ \lambda_X } \frac{  -\ii \lambda_X -m_0}{-\ii \lambda_X +m_0},\label{eq:detratio}
\eeq
where the product runs over all eigenvalues $\lambda_X$ of the Dirac operator $\CD_X$.
We define $s(\lambda_X)$ by
\beq
e^{-\pi \ii s(\lambda_X)} = \frac{  -\ii \lambda_X -m_0}{-\ii \lambda_X +m_0},~~~ -1< s(\lambda_X) \leq 1.
\eeq
This $s(\lambda_X)$ is essentially $\sign(\lambda_X)=\lambda_X/ |\lambda_X|$ for a large $m_0$. 
But it has the properties that $s(\lambda_X) \to 0 $ for $\lambda_X \to \pm \infty$ and $s(0)=+1$.
Therefore, $\sum_{\lambda_X} s(\lambda_X)$ can be considered as a regularized version of $\sum_{\lambda_X \neq 0} \sign(\lambda_X) +\dim {\rm Ker}\CD_X$,
which in turn is the $2 \eta(T)$.\footnote{In quantum field theory, it is believed that different regularizations give the same answer up to local counterterms.
In the current problem, there seems to be no candidates for a counterterm which could affect our results. Therefore the regularization here
is expected to give the same answer as the usual zeta regularization.} Thus,
\beq
\lim_{m_0 \to \infty} \frac{\bra{T} \et{X}_{ m= - m_0}}{\bra{T} \et{X}_{ m= + m_0}} &= \lim_{m_0 \to \infty} \exp( -\ii \pi \sum_{\lambda_X} s(\lambda_X) ) \nonumber \\
&= e^{-2\pi \ii \eta(T)}. \label{eq:rat1}
\eeq

Next, we rewrite $\bra{T} \et{X}$ by using the ground state $\ket{\Omega}$. For this purpose, let us note the following point.
In the above discussion, we assumed that $X$ has the cylindrical boundary region and $X= X' \cup [-\tau_0, 0] \times Y$.
This means that the amplitude is given as
\beq
\bra{T} \et{X} = \bra{T} e^{-\tau_0 H} \ket{X'}
\eeq
where $ \ket{X'}$ is the state created by the path integral on $X'$.
All the states other than the ground state have energies larger than or equal to $ |m|$. If we take the limit $|m| \to \infty$,
the factor $e^{-\tau_0 H}$ projects out all the states other than the ground state $\ket{\Omega}$. Therefore,
we get $ \bra{T} e^{-\tau_0 H} \ket{X'} \to  \bra{T}\et{\Omega}  \bra{\Omega} e^{-\tau_0 H} \ket{X'}$ for $|m| \to \infty$
and hence
\beq
\lim_{m_0 \to \infty} \frac{\bra{T} \et{X}_{ m= - m_0}}{\bra{T} \et{X}_{ m= + m_0}}  &= \lim_{m_0 \to \infty}
\frac{  \bra{T}  \et{\Omega_{-m_0}}   \bra{\Omega_{-m_0}} \et {X}_{ m= - m_0}  }{  \bra{T}  \et{\Omega_{m_0}}   \bra{\Omega_{m_0}} \et {X}_{ m= m_0} } \nonumber \\
&=\det(T) \lim_{m_0 \to \infty}
\frac{   \bra{\Omega_{-m_0}} \et {X}_{ m= - m_0}  }{   \bra{\Omega_{m_0}} \et {X}_{ m= m_0} }   \label{eq:rat2}
\eeq
where we have used \eqref{eq:ratiodetT}. From \eqref{eq:rat1} and \eqref{eq:rat2}, we see that
\beq
e^{-2\pi \ii \eta(T_2)}=\det(T_2 T^{-1}_1) e^{-2\pi \ii \eta(T_1)}.\label{eq:DF1}
\eeq
This is the formula \eqref{eq:DFth1} of Theorem~1.

\subsection{Derivation of Theorem 2}
Next let us derive the gluing formula \eqref{eq:DFth2}.
Let $Z$ be a codimension one subspace of $X$ such that the neighborhood of $Z$ in $X$ is given by a cylinder $(-\tau_0, \tau_0) \times Z$.  Then
let $X^{\rm cut}$ be a manifold which is obtained by
cutting $X$ along Z. If $\partial X = Y$, we have $\partial X^{\rm cut} = Y \sqcup Z \sqcup -Z$.

The path integral over $X^{\rm cut}$ produces an element of the Hilbert space 
$\CH_{\partial X^{\rm cut} } \cong \CH_Y \otimes \CH_{Z \sqcup -Z} $.
We denote this state as $\ket{X^{\rm cut}}$. A generalized APS boundary condition $T_Y \oplus T_{Z \sqcup -Z}$ is imposed as described in the Introduction.

There are natural isomorphisms
\beq
\CH_{Z \sqcup -Z} \cong  \CH_{-Z} \otimes \CH_Z  \cong \CH_{Z}^* \otimes \CH_Z \label{eq:iso1}
\eeq
where $\CH_Z^*$ is the dual space to $\CH_Z$. There is also a natural map 
\beq
 \CH_{Z}^* \otimes \CH_Z  \to \bC, \label{eq:contraction}
\eeq
which is defined by $\bra{\alpha} \otimes \ket{\beta} \mapsto \bra{\alpha} \et{\beta}$.\footnote{
Here we are carefully distinguishing $\bra{\alpha} \otimes \ket{\beta} $ from $  \ket{\beta} \otimes \bra{\alpha}$.
There is an isomorphism $\CH_{Z}^* \otimes \CH_Z \cong  \CH_Z \otimes \CH_{Z}^* $, 
but under this isomorphism, the state $\bra{\alpha} \otimes \ket{\beta} $ goes to the state $ (-1)^{F_\alpha F_\beta} \ket{\beta} \otimes \bra{\alpha}$,
where $F_\alpha $ and $ F_\beta $ are 0 or 1 depending on the bose-fermi statistics of $\ket{\alpha}$ and $\ket{\beta}$, respectively. 
For example, this factor is responsible for the anti-periodic boundary condition in the thermal partition function of a fermion. 
Restricting our attention to the ground state, this leads to a grading of the line bundle in which the ground state takes values.
This grading should corresponds to the grading discussed in \cite{Dai:1994kq}. In \cite{Dai:1994kq}, 
the grading gave an additional sign factor $(-1)^{{\rm Ind}(\CD_Z)}$ in the formula \eqref{eq:DFth2},
but we don't have that factor probably because of a slight difference of our convention for the APS boundary conditions. }
The path integral (or more generally the axioms of quantum field theory) tells us that under the composition of these maps,
the state $\ket{X^{\rm cut}}$ maps to the state $\ket{X}$. Furthermore, because of the Euclidean time evolution $e^{-H \tau_0}$ 
in the cylindrical region $(-\tau_0, \tau_0) \times Z$,
these states are all proportional to the ground states of the respective Hilbert spaces in the limit $|m| \to \infty$. Therefore, in this limit we obtain
\beq
\ket{X^{\rm cut}} \to \ket{X} \otimes \bra{\Omega(Z)} \otimes \ket{\Omega (Z)}.
\eeq
where $\ket{\Omega(Z)}$ is the ground state on $Z$.

We assume that under the isomorphism $ \CH_{Z}^* \otimes \CH_Z   \cong \CH_{Z \sqcup -Z}$,
the state $\bra{\Omega(Z)} \otimes \ket{\Omega (Z)}$ maps to
\beq
\bra{\Omega(Z)} \otimes \ket{\Omega (Z)} \mapsto \xi(Z) \ket{\Omega(Z \sqcup -Z)}
\eeq
where $\ket{\Omega(Z \sqcup -Z)}$ is the ground state determined by the procedure discussed in Sec.~\ref{sec:boundary},
and $\xi(Z)$ is a phase factor which only depends on $Z$ and the mass $m$. 
Both the left and right hand side are the ground states, so there is only a phase ambiguity represented by $\xi(Z)$ in this correspondence.
Thus we learned that 
\beq
\ket{X^{\rm cut}} \to \ket{X} \otimes \xi(Z) \ket{\Omega(Z \sqcup -Z)}
\eeq
in the large mass limit.

At this point, we can simply use the formula \eqref{eq:rat2} for $X^{\rm cut}$ to get
\beq
&\lim_{m_0 \to \infty} \frac{\bra{T_Y\oplus T_{Z \sqcup -Z}} \et{X^{\rm cut}}_{ m= - m_0}}{\bra{T_Y\oplus T_{Z \sqcup -Z}} \et{X^{\rm cut}}_{ m= + m_0}}  \nonumber \\
=&\det(T_Y) \det(T_{Z \sqcup -Z})  \lim_{m_0 \to \infty} 
\frac{   \xi(Z)|_{ m= -m_0}    }{  \xi(Z)|_{ m= m_0}      } 
\frac{   \bra{\Omega_{-m_0}(Y) } \et {X}_{ m= - m_0}  }{   \bra{\Omega_{m_0} (Y)} \et {X}_{ m= m_0} } .
\eeq
Again using \eqref{eq:rat2} for $X$, we get the exponentiated eta invariant $e^{-2\pi \ii \eta_{X^{\rm cut}}(T_Y\oplus T_{Z \sqcup -Z})}$ for $X^{\rm cut}$ as
\beq
e^{-2\pi \ii \eta_{X^{\rm cut}}(T_Y\oplus T_{Z \sqcup -Z})}=  \zeta(Z) \det(T_{Z \sqcup -Z}) e^{-2\pi \ii \eta_X(T_Y)}.
\eeq
where $\zeta(Z)= \lim_{m_0 \to \infty}  ( \xi(Z)|_{ m= -m_0}    /  \xi(Z)|_{ m= m_0}  ) $. This phase factor $\zeta(Z)$ only depends on $Z$.

In principle, $\zeta(Z)$ can be determined by a careful examination of the isomorphism $ \CH_{Z}^* \otimes \CH_Z   \cong \CH_{Z \sqcup -Z}$.
Instead of doing that, we will determine it by considering a simple example of $X$ and $X^{\rm cut}$. 
However, before doing that, we remark that the precise (basis-independent) meaning of 
$\det(T_{Z \sqcup -Z}) $ is given by using the isomorphism $H_+(Z \sqcup -Z) \cong H_+(Z) \oplus H_+(-Z) \cong H_-(-Z) \oplus H_-(Z) \cong H_-(Z \sqcup -Z) $
as in the Introduction. Then $T_{Z \sqcup -Z}$ can be regarded as an endomorphism and the determinant is well-defined.
For example, one can check that the standard APS boundary condition is given by
\beq
U_{Z \sqcup -Z} = \left( \begin{array}{cc}
0 & -U_Z^\dagger \\
U_Z & 0
\end{array} \right).\label{eq:sAPS2}
\eeq
where $U_Z$ is defined in \eqref{eq:sAPS}.
The appearance of $-U_Z^\dagger$ is due to the fact that $\CD_X = \ii \gamma^\tau( \partial_\tau +\CD_Y) =  \ii \gamma^{-\tau}( \partial_{-\tau} -\CD_Y)$ 
and hence $\CD_{-Y}=-\CD_Y$, and taking into account chirality we get $\CD^{-+}_{-Y}=-\CD_Y^{+-}$.
In this case, we get $\det(U_{Z \sqcup -Z} )=1$.
On the other hand, our computation in Sec.~\ref{sec:boundary} 
did not take into account this natural isomorphism $H_+(Z \sqcup -Z)  \cong H_-(Z \sqcup -Z) $ and 
in particular the $\det(T)$ in the formula \eqref{eq:ratiodetT} was taken with respect to an arbitrarily chosen basis.
We absorb this phase ambiguity into $\zeta(Z)$.

Now let us determine $\zeta(Z)$.
We take $X = S^1 \times Z$ and $X^{\rm cut} = [0,1] \times Z$.
One can perform Kaluza-Klein decomposition on $Z$ and reduce the problem to a one-dimensional problem where $X=S^1$ and $X^{\rm cut}=[0,1]$.
Furthermore, one can check that nonzero modes of the Kaluza-Klein decomposition do not contribute to the eta-invariant 
if we impose the standard APS boundary conditions for these nonzero modes. This is because eigenvalues of the same absolute value with positive and negative signs
always appear in pairs for these modes. Therefore, we only need to care about zero modes on $Z$. (This can be still nonzero modes on the direction $S^1$.)

Let us concentrate on zero modes on $Z$, which we write as $\Psi_0$. Furthermore, 
we take the boundary condition as $T_{Z \sqcup -Z} |_0=1$, where $|_0$ means the restriction to the zero modes.
We need to carefully examine what this means. Our boundary condition was that $\vec{B}_1=0$ where $\vec{B}_1$ was defined in \eqref{eq:APSmode}.
However, the $\vec{A}_+$ and $\vec{A}_-$ in that equation are actually given by 
\beq
\vec{A}_+ = ( \Psi_{+,0}(\tau=1),  \Psi_{-,0}(\tau=0)),~~~~~\vec{A}_- = ( \Psi_{+,0}(\tau=0),  \Psi_{-,0}(\tau=1)).
\eeq
Therefore, under the isomorphism $H_+(Z \sqcup -Z)  \cong H_-(Z \sqcup -Z) $,
the condition $T_{Z \sqcup -Z} |_0=1$ means that $\Psi_0(\tau=1)=\Psi_0(\tau=0)$ 
and hence the $\Psi_0$ behaves just as if they are living on the original $X=S^1$ before the cutting. Therefore, for this choice of $T_{Z \sqcup -Z}$,
we get 
$
e^{-2\pi \ii \eta_{X^{\rm cut}}( T_{Z \sqcup -Z})}=e^{-2\pi \ii \eta_X}
$
because both the left-hand-side and right-hand-side are computed in completely the same way.
On the other hand, for the above choice of $T_{Z \sqcup -Z}$, we also get $\det(T_{Z \sqcup -Z})=1$.
Because $\zeta(Z)$ is independent of the $T_{Z \sqcup -Z}$, we conclude that $\zeta(Z)=1$. 

In summary, we get
\beq
e^{-2\pi \ii \eta_{X^{\rm cut}}(T_Y\oplus T_{Z \sqcup -Z})}=  \det(T_{Z \sqcup -Z}) e^{-2\pi \ii \eta_X(T_Y)}. \label{eq:DF2}
\eeq
This is the gluing law of the exponentiated eta-invariant \eqref{eq:DFth2}.
For the standard APS boundary condition \eqref{eq:sAPS2}, we have $ \det(U_{Z \sqcup -Z})=1$ and the gluing law has a simple form.


\section{The ground state and Berry phase}\label{sec:Berry}
If we integrate out the massive fermion, the theory seems to be ``empty" whose Hilbert space is one-dimensional and is spanned by the ground state of the massive fermion.
Nethertheless, this one-dimensional Hilbert space can be nontrivial.\footnote{Topological field theories whose Hilbert spaces on any manifolds without boundary are one-dimensional
are called invertible field theories~\cite{Freed:2014iua,Freed:2014eja,Freed:2016rqq}.} 
The purpose of this section is to discuss this nontrivial behavior of the ground state.

In this section, we assume that the theory with positive mass parameter $m>0$ gives a trivial ground state,
and we consider the theory with negative mass parameter $m<0$. 
Then we omit to take ratios of amplitudes of these two theories as we did in the previous sections. 
If one prefers it, one might think of taking the ratios to be just a Pauli-Villars regularization, regarding the positive $m$ theory as the Pauli-Villars regulator. 

\subsection{Berry phase}\label{sec:Bphase}
In a quantum system with a large mass gap, we can consider an adiabatic process of changing parameters such as
the shape of a material (or metric in our case) 
and the external electromagnetic field (or background gauge field in our case) in such a way that the system remains to be in the ground state.
When we go through such an adiabatic process and return to the same point in the parameter space, the ground state $\ket{\Omega}$ may acquire a phase factor
\beq
\ket{\Omega,{\rm final}} = e^{\ii \CB} \ket{\Omega,{\rm initial}}.
\eeq
This $\CB$ is the Berry phase.

In our context of the massive fermion theory, the Berry phase can be computed as follows.
We put the theory on a compact $d$-dimensional space $Y$ with background field, and change the parameters as time evolves.
Let $W$ be a parameter space of metrics and background gauge fields. Each point $w$ on $W$ specifies a metric and gauge field on $Y$,
and we denote the manifold $Y$ equipped with that metric and gauge field as $Y_w$.
In this space $W$, we consider a path $\gamma(\tau)~(0 \leq \tau \leq \tau_1)$ from one point $w=\gamma(0) \in W$ to another $w'=\gamma(\tau_1) \in W$.
Now we regard $\tau$ as the (Euclidean)\footnote{For the ground state, the difference between Euclidean and Lorentzian time evolutions does not matter
and only the adiabatic change of the state vector is important.} time, and
we take the spatial components of the metric $g$ and the background gauge field $A$ at the time $\tau$ to be the one specified by $\gamma(\tau) \in W$.
This process defines a $d+1$ dimensional manifold $Y_\gamma$ which is topologically $  [0,\tau_1] \times Y$.
Then we define a parallel transport of the ground state $\ket{\Omega}$ from $w$ to $w'$ along $\gamma$
by the Euclidean path integral. We denote the situation as
\beq
|Y_\gamma|\et{\Omega},
\eeq
where the notation $|Y_\gamma|$ means the path integral on $Y_\gamma$ which gives a map from the Hilbert space $\CH_{Y_w}$ at the point $w$ 
to $\CH_{Y_{w'}}$ at $w'$,
\beq
|Y_\gamma| : \CH_{Y_w} \to \CH_{Y_{w'}}.
\eeq
In this way, we can define a parallel transport of the ground state in the space $W$.

Next we consider the case that the path $\gamma$ forms a loop $\gamma(0)=\gamma(\tau_1)$ inside $W$.
This means that the metric and gauge field at $\tau=0$ and $\tau=\tau_0$ are the same up to a diffeomorphism and gauge transformation.
The manifold $Y_\gamma$ is a fiber bundle with the fiber $Y$ and the base $S^1$.
The parallel transport of the ground state gives 
the Berry phase $ e^{\ii \CB} = \bra{\Omega,{\rm initial}}\et{\Omega,{\rm final}}= \tr | Y_\gamma |$.
We denote the Berry phase along the path $\gamma$ as $\CB(\gamma)$.
The parallel transport is defined by the path integral, and the path integral gives the exponentiated eta-invariant, so we get
\beq
e^{\ii \CB(\gamma)} = \exp({-2 \pi \ii \eta_{Y_\gamma}}). \label{eq:Beta}
\eeq
This is the formula for the Berry phase.

\subsection{Berry connection and curvature}\label{sec:conn}
Let us slightly rephrase the above situation. 
At each point of the parameter space $W$, we have $Y_w$ with the specific metric and background gauge field. 
Then, this defines a fiber bundle $F$ 
\beq
 \pi: F \to W,~~~~~\pi^{-1}(w) = Y_w.
\eeq
The base is $W$ and the typical fiber is $Y$. We assume that the metric and the gauge field are extended into the total space $F$. In particular the metric is 
\beq
 ds^2=g_{ij}(y,w) (dy^i - B^i_a(y,w) d w^a)(dy^j - B^j_b(y,w) d w^b)+\frac{1}{\epsilon^2}g_{ab}(w)dw^a dw^b.
\eeq
Then, a path $\gamma$ in the base $W$ can be lifted to a manifold $Y_\gamma = \pi^{-1}(\gamma)$ in the total space $F$.
Adiabaticity (i.e., slow change of the metric and gauge field on $Y$) is achieved by taking $\epsilon \to 0$.

Now we rewrite the Berry phase. The exponentiated eta-invariant has the property that it is given by the exponentiated Chern-Simons invariant up to a constant phase.
The Chern-Simons invariant which is relevant to our fermion is given by $\int I^0_{d+1}$, where $ I^0_{d+1}$ is a $d+1$-form characterized by
\beq
dI^0_{d+1} &=I_{d+2}= \left. \hat{A}(R) \tr \exp( \frac{\ii F}{2\pi}) \right|_{d+2},
\eeq
where $\hat{A}(R) $ is the $\hat{A}$ genus of the metric and $F$ is the curvature of the background gauge field.
Then, we have
\beq
 \exp\left(-2\pi \ii \eta_{Y_\gamma}  \right) = [{\rm const.}]  \exp\left(2\pi \ii \int_{Y_\gamma}  I^0_{d+1}  \right).    \label{eq:invcomb}
\eeq
When the manifold $Y_\gamma$ is an even dimensional $\pin^\pm$ manifold, the Chern-Simons 
invariant is defined to be zero and hence $\exp({-2 \pi \ii \eta_{Y_\gamma}})$ is a topological invariant given by $[{\rm const.}]$.

Strictly speaking, the Chern-Simons invariant should be defined in more gauge invariant way. For example, if the 
$\gamma$ is a boundary of a disk $D_\gamma \subset W$, then we can define
\beq
\int_{Y_\gamma}  I^0_{d+1} := \int_{\pi^{-1} (D_\gamma)}  I_{d+2}.\label{eq:diskf}
\eeq
More generally, the $\gamma$ can be topologically nontrivial inside $W$. However, we may take a reference path $\gamma_0$ 
and the homotopy $\gamma_s(\tau)=\Gamma(\tau,s)~~(0 \leq s \leq 1)$ such that $\gamma_1=\Gamma(\cdot, 1)=\gamma$ and $\gamma_0=\Gamma(\cdot, 0)$. Then,
the Chern-Simons invariant on $Y_\gamma$ may be defined up to constant by $\int_{\Gamma^* F} \Gamma^* I_{d+2}$.
In this case, the $[{\rm const.}]  $ in \eqref{eq:invcomb} is given in terms of the reference path $\gamma_0$ by $ \exp\left(-2\pi \ii \eta_{Y_{\gamma_0}}  \right)$.
Whenever we write an expression like $\int_{Y_\gamma}  I^0_{d+1}  $, one may interpret it in this way.
But in the following, we proceed as if $I^0_{d+1}$ is well-defined for simplicity because it might be more intuitive.

The formula \eqref{eq:invcomb} is a consequence of the APS index theorem~\cite{Atiyah:1975jf}. Another physical explanation is as follows. 
The $\exp({-2 \pi \ii \eta_{X}})$ appeared as the partition function of the massive fermion on $X$.
However, the low energy effective action
after integrating out the massive fermion is, at least locally, given by $2 \pi \ii  \int_X I^0_{d+1}$ by a one-loop Feynman diagram (or whatever\footnote{
A simple method of computation which just uses the usual Atiyah-Singer index theorem is as follows. On a manifold of the form $X=\bR \times Y$,
the Chern-Simons term can be written schematically as $ \int A_{0}d\tau \wedge F^{ \frac{d}{2}}$ where for simplicity we only considered gauge field.
In this form, the Chern-Simons term can be interpreted as giving the charge of the ground state under nontrivial instanton numbers of $ F^{ \frac{d}{2}}$.
In the UV fermion, that charge can be computed simply by quantizing the zero modes of the fermion in instanton backgrounds.
See e.g., Sec.~2.2 of \cite{Ohmori:2014kda} for this procedure.
}) computation. Thus
$\exp({-2 \pi \ii \eta_{X}})$ must be equal to $\exp(2 \pi \ii  \int_X I^0_{d+1})$ up to a factor that is invariant under local deformation.
Another derivation of \eqref{eq:invcomb} is given in Appendix~\ref{sec:convention}, where we also fix our conventions for gamma matrices.

Using the interpretation of $Y_\gamma$ as a subspace of the total space of the fiber bundle $F$, 
the Chern-Simons invariant can be rewritten as follows. Let us consider the integral over a fiber $Y_w$
\beq
\CA  = -2 \pi \ii  \int_{Y_w}  I^0_{d+1}.
\eeq
This is integrating the $d+1$ form $I^0_{d+1}$ on a $d$-dimensional space $Y_w$, so it locally defines a one-form $\CA$ on the base parameter space $W$.
However, this $\CA$ is not globally defined as a 1-form over the whole $W$, and
globally $\CA$ defines a connection of a certain line bundle which will be discussed in the next subsection.
This is the connection associated to the parallel transport discussed above.
We get the Chern-Simons invariant as
\beq
2\pi \ii  \int_{X_\gamma}  I^0_{d+1} =- \int_\gamma \CA.
\eeq

Therefore, the Berry phase may be written as a holonomy
\beq
e^{\ii \CB(\gamma)} = \exp\left( - \int_\gamma \CA \right),  \label{eq:Bconn}
\eeq
where our convention for the gauge field and connections is such that they are anti-hermitian and hence $\CA$ is pure imaginary.
Here we did not write the factor $[{\rm const.}]$ in \eqref{eq:invcomb}, which is very important for the case $I_{d+2}=0$, such as $\pin^\pm$ manifolds with odd $d$.
The reason for not writing this factor is that we are including this factor into the definition of the holonomy $\exp\left( - \int_\gamma \CA \right)$.
Locally in the $\pin^\pm$ case, the connection $\CA$ defined above is zero, but globally it can have a nontrivial structure 
which takes into account the factor $[{\rm const.}]$.

The formula \eqref{eq:Bconn} immediately implies that $\CA$ is the Berry connection.
Denoting the exterior derivative on $W$, $Y$ and $F$ as $d_w$, $d_y$ and $d$ respectively, where $d=d_w+d_y$, 
the Berry curvature $\CF=d_w \CA$ is computed as
\beq
\CF&=d_w \CA =-2 \pi \ii  \int_{Y_w}  (d-d_y)I^0_{d+1} \nonumber \\
&=-2 \pi \ii  \int_{Y_w}  I_{d+2} \label{eq:curva}
\eeq
where we used integration by parts $ \int_{Y_w} d_yI^0_{d+1} =0$ and also used $dI^0_{d+1}=I_{d+2}$.
(The formula \eqref{eq:curva} actually follows directly from \eqref{eq:diskf}.)
More explicitly, we get
\beq
\CF=   -2 \pi \ii \int_{Y_w}  \left. \hat{A}(R) \tr \exp( \frac{\ii F}{2\pi}) \right|_{d+2} \label{eq:Bcurvature}
\eeq
This is the formula for the Berry curvature. We are integrating the $d+2$-form on the $d$-dimensional manifold $Y_w$,
so we get a 2-form $\CF$ on $W$. Explicit examples are given in Sec.~\ref{sec:example}.

\subsection{The ground state and the determinant line bundle}\label{sec:detline}

The ground state spans a complex one-dimensional vector space $\CL_w \subset \CH_{Y_w}$ over each point $w$ of the parameter space $W$.
This defines a line bundle $\CL$ over the space $W$. The parallel transport, connection and curvature defined above are the ones on this line bundle.
We argue that this bundle is what is called the determinant line bundle associated to the Dirac operator $\CD_Y$.

First, notice the following point. The states $\ket{T}$ defined by generalized APS boundary conditions are unambiguously defined including their
phase factors (at least if we take the ratio of the theories with $m<0$ and $m>0$). 
The reason is that these states are just defined by generalized APS boundary conditions in the path integral, and also they are independent of the mass parameter $m$.
There is simply no room for phase ambiguity to arise in the computation of the path integral with the boundary condition $T$; see \eqref{eq:detratio}.
Notice also that the projector $\ket{\Omega}\bra{\Omega}$ to the ground state is also uniquely determined without
any phase ambiguity. Therefore, the state vector
\beq
\ket{\Omega,T} :=  \ket{\Omega}\bra{\Omega}\et{T}
\eeq
is unambiguously defined (at least if we take the ratio for $m<0$ and $m>0$).

The state $\ket{\Omega, T}$ depends on the choice of $T$. However, the standard APS boundary condition $U_Y$ given in \eqref{eq:sAPS}
is uniquely defined. If $U_Y$ were well-defined everywhere in the parameter space $W$, then
the $\ket{\Omega,U_Y}$ would have given a global section of the line bundle $\CL$ without zero and hence $\CL$ could have been trivialized.
However, when some of the eigenvalues of $\CD_Y$ become zero, the standard APS boundary condition becomes ill-defined
and this is the obstruction for trivializing the line bundle $\CL$. In general, there is no globally well-defined boundary condition
over the whole parameter space $W$. For example, the first Chern class of the line bundle is given by
\beq
c_1(\CF)=\frac{\ii \CF}{2\pi}=  \int_{Y_w}  \left. \hat{A}(R) \tr \exp( \frac{\ii F}{2\pi}) \right|_{d+2}
\eeq
and this can give nontrivial values when integrated over a two-cycle on $W$ as we will discuss in an explicit example in the next section.

The best thing we can do is the following. We take open covering $\{V_\alpha \}$ of the parameter space $W$ so that $W=\bigcup_\alpha V_\alpha$.
Each of $V_\alpha$ is assumed to be small enough so that we can pick up a single boundary condition $T_\alpha$ which is defined over $V_\alpha$.
(More precisely it is a local section $T_\alpha(w)~( w \in V_\alpha)$ of the bundle with fiber $H_+(Y_w)^* \otimes H_-(Y_w)$ which varies smoothly over $V_\alpha$.)
Then we can take a local section of $\CL$ given by $\ket{\Omega,T_\alpha}$ in each $V_\alpha$.
If $V_\alpha \cap V_\beta \neq \varnothing$, the transition function is given by
\beq
\ket{\Omega,T_\beta}=\det(T^{-1}_\beta T_\alpha)\ket{\Omega,T_\beta}.
\eeq
This follows from \eqref{eq:ratiodetT}. Therefore, the line bundle $\CL$ is defined by these transition functions $\det(T^{-1}_\beta T_\alpha)$.
This line bundle is the determinant line bundle of $\CD_Y$, as will discuss below.
In this way, we found that the ground state takes values in the determinant line bundle.
The natural connection and curvature of this line bundle are the Berry connection and curvature constructed in the previous subsection.
The relation to the line bundle discussed in the Introduction will be made more explicit in Sec.~\ref{sec:Th3}.

Let us explain the reason why the above line bundle is called the determinant line bundle of $\CD_Y$.
(Depending on convention, it was called the inverse of the determinant line bundle in \cite{Dai:1994kq}.)
If we consider a $d$-dimensional massless chiral fermion with negative chirality (as opposed to $d+1$ dimensional massive fermion),
the path integral gives the partition function as
\beq
 \det( \CD^{+-}_Y).
\eeq
where $\CD^{+-}_Y$ is the Dirac operator $\CD_Y$ acting on the negative chirality fermion in a $d$-dimensional manifold $Y$.
However, this $\CD^{+-}_Y$ is a map from the space of sections $H_-(Y)$ with negative chirality to the space of sections $H_+(Y)$ with positive chirality.
Because the vector spaces $H_+(Y)$ and $H_-(Y)$ are not naturally isomorphic, the expression $ \det( \CD^{+-}_Y)$ does not naturally give a numerical number in $\bC$.
Rather, it takes values in the one dimensional vector space $\det H_+(Y) \otimes \det H_-(Y)^*$.
One way to get a number in $\bC$ it may be to pick up a unitary map $T$ from $H_+(Y)$ to $H_-(Y)$, and consider 
\beq
\det( \CD^{+-}_Y T ).
\eeq
This is now the determinant of a map from $H_+(Y)$ to $H_+(Y)$, and hence gives a value in $\bC$ (after a suitable regularization).
However, it now depends on the choice of $T$. Under a change of $T$ it behaves as
\beq
\det( \CD^{+-}_Y T_\beta )=\det(T_\beta T_\alpha^{-1}) \det( \CD^{+-}_Y T_\alpha ).
\eeq
This is precisely the opposite transformation law as that of $\ket{\Omega,T}$. Therefore, over the parameter space $W$,
there is a section of the line bundle $\CL$ given by
\beq
\ket{\Omega,T}\det( \CD^{+-}_Y T ).
\eeq
Because of this, $\det( \CD^{+-}_Y )$ can be regarded as a section of the line bundle $\CL$. In general, this global section has zero at some points of $W$,
and that is the obstruction for trivializing $\CL$.

We remark that even if the line bundle $\CL$ can be trivialized, the low energy theory can be still nontrivial.
The argument for this is a ``wick rotation" of the argument in \cite{Witten:2015aba,Witten:2016cio}.
If we are given a $d+1$ manifold $X$ with $\partial X=Y$, we get a state vector $\ket{X} \in \CH_Y$ by the path integral on $X$.
By the Euclidean time evolution $e^{- \tau_0 H}$, it is proportional to the ground state in the large mass limit.
If $\ket{X}$ does not depend on $X$ at all, then the low energy theory is really trivial. If it depends on $X$, that means
that when we take another manifold $X'$ with $\partial X'=Y$, then $\bra{X'}\et{X}$ gives a nontrivial value.
Such a nontrivial amplitude gives a nontrivial low energy topological field theory.

\subsection{Derivation of Theorem 3}\label{sec:Th3}
We can now prove Theorem~3 of the Introduction. 
The first question is what is $\CT$ defined in \eqref{eq:tausection} in our context. Using the results of Sec.~\ref{sec:DF}, we can write it as
\beq
\CT(X) = \frac{\bra{T}\et{X} }{ \det T} = \frac{ \bra{T}\et{\Omega} \bra{\Omega} \et{X} }{\det T}= \frac{ \bra{\Omega,T} \et{X} }{\det T}
\eeq
where the large mass limit $|m| \to \infty$ is implicit. The combination
\beq
\frac{\bra{\Omega,T}}{\det T} \label{eq:iso}
\eeq
is independent of $T$ due to \eqref{eq:ratiodetT}. The important point is that \eqref{eq:iso} gives the natural isomorphism between the determinant line bundle $\CL$
defined in the Introduction and the bundle of the ground state $\CL$ defined in this section; Anticipating this isomorphism, we have already used the same notation 
$\CL$ for both of the line bundles. Therefore, under this natural isomorphism, we identify
\beq
\CT(X)  = \ket{X}.
\eeq

In Sec.~\ref{sec:Bphase} we have already defined a parallel transport of vectors in $\CL$
from $w \in W$ to $w' \in W$. This gives a connection $\nabla$ by the standard argument. This is the connection which we have discussed in Sec.~\ref{sec:conn}.

\begin{figure}
\centering
\begin{tabular}{c@{\qquad}c}
\includegraphics[width=.7\textwidth]{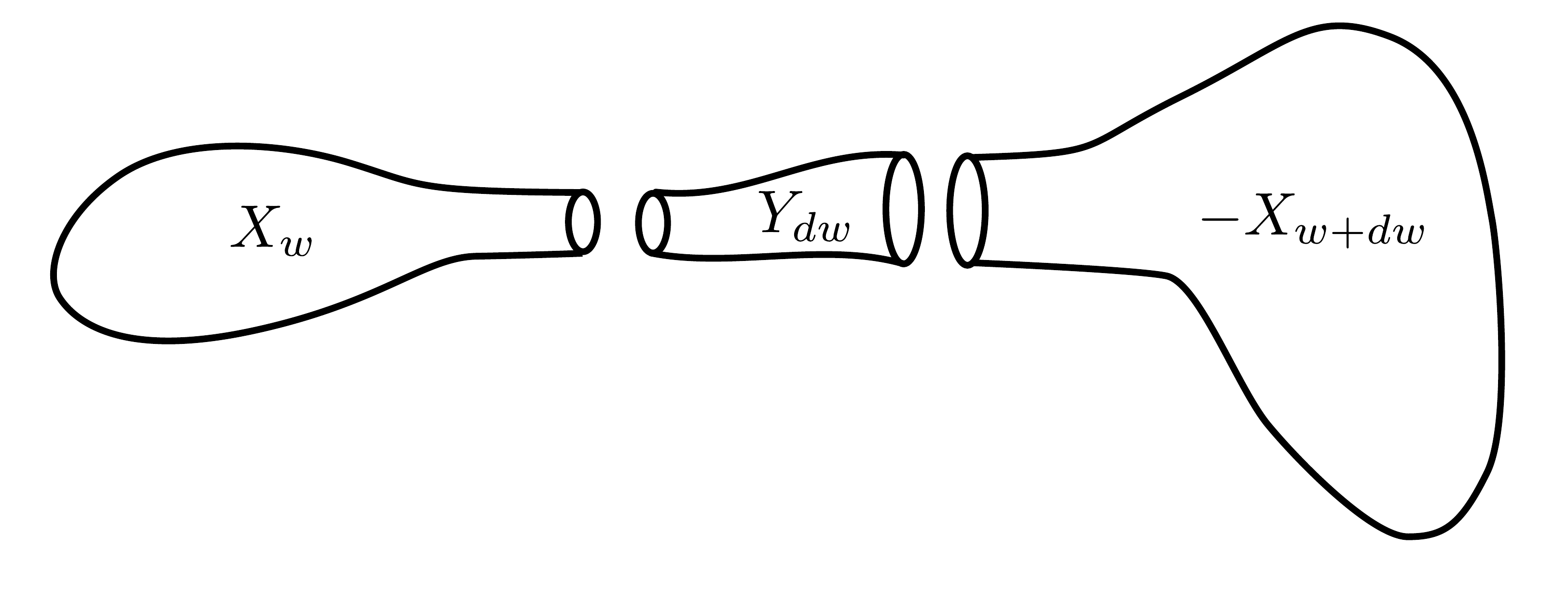}\\
\end{tabular}
\caption{The manifolds $X_w$, $-X_{w+dw}$ and the parallel transport given by the manifold $Y_{dw}$. They glue together to give a manifold without boundary.
\label{fig:zu2}}
\end{figure}

The remaining task is to derive \eqref{eq:DFth3}. Up to now, we have considered the fiber bundle $F \to W$ whose fiber is $Y$.
Now we consider the bundle where the fiber at $w \in W$ is $X_w$ with $\partial X_w=Y_w$, as discussed in the Introduction.
We want to compare two state vectors $\ket{X_w}$ and $\ket{X_{w+dw}}$ for infinitesimally close points $w$ and $w+dw$.
Their boundaries are $Y_w =\partial X_w$ and $Y_{w+dw} = \partial X_{w+dw}$, respectively. To compare them,
we need to do parallel transport from $Y_w$ to $Y_{w+dw}$. We denote the manifold interpolating them as $Y_{dw}$,
where $\partial Y_{dw} = Y_{w+dw} -Y_{w}  $. See Fig.~\ref{fig:zu2} for the situation.
Then the covariant exterior derivative is given by
\beq
\nabla  \CT (X_w) =\nabla \ket{X_w}=  \ket{X_{w+dw}} - |Y_{dw} \ket{X_w} 
\eeq
or more conveniently
\beq
\nabla  \log \CT (X_w) & =  1- \bra{X_{w+dw}} Y_{dw} \ket{X_w} \nonumber \\
& = -\log \left( \bra{X_{w+dw}} Y_{dw} \ket{X_w} \right).
\eeq
The value of $\bra{X_{w+dw}} Y_{dw} \ket{X_w}$ is given by the exponentiated eta-invariant on the manifold ${X}_{\rm com}$
which is constructed by gluing $X_w$, $Y_{dw}$ and $-X_{w+dw}$. See Fig.~\ref{fig:zu2}.
Here the meaning of the minus sign in $-X_{w+dw}$ is that 
we take the ``reflection of $X_{w+dw}$ in the time direction"
in the way which is used to formulate the reflection positivity in Euclidean quantum field theory. 
The procedure is that we first take the reflection in the cylindrical region $(-\tau_0, 0] \times Y_{w+dw}$ in the direction $\tau$,
and then extend the structure to the entire $X_{w+dw}$.
For our purposes here, we just need the fact that the orientation is flipped in the cases of orientable manifolds. For unorientable cases, we can just take $I^0_{d+1}=0$
in the following computation and hence the details of the reflection does not matter.
More complete treatment of the reflection $-X$ may be found in \cite{Freed:2016rqq}.

The eta-invariant is given by integration of $I^0_{d+1}$ on ${X}_{\rm com}$.
This in turn is given by the sum of integrations over $X_w$, $-X_{w+dw}$ and $Y_{dw}$.
We get
\beq
 &-\log \left( \bra{X_{w+dw}} Y_{dw} \ket{X_w} \right) \nonumber \\
 =& \left( 2\pi \ii \int_{X_{w+dw}} I^0_{d+1} \right) - \left(   2\pi \ii \int_{X_{w}} I^0_{d+1} \right) -(-\CA) \nonumber \\
 =& 2\pi \ii d_w \int_{X_{w}}  I^0_{d+1} -2 \pi \ii  \int_{Y_w}  I^0_{d+1} \nonumber \\
  =& 2\pi \ii \int_{X_{w}} d_w I^0_{d+1} +2 \pi \ii  \int_{X_w} d_x I^0_{d+1} \nonumber 
\eeq
where $x$ are the coordinates of $X$, $d_w =dw \cdot \partial_w$ and $d_x=dx \cdot \partial_x$.
The term $\CA$ comes from the parallel transport along $Y_{dw}$, because $\CA$ is the connection of the parallel transport. 
The change of the sign of the term $2 \pi \ii  \int_{X_w} d_x I^0_{d+1}$ in the last equality requires a care. 
The precise meaning of integrating a $d+2$-form $\alpha_{d+2}$ on $X$ is that we first write it as $\alpha_{d+2} = dw \wedge (A \ d^{d+1}x)+\cdots$
and define $\int_X \omega_{d+2} := dw \int_X A d^{d+1}x $. In the same way, integrating a $d+1$-form $\beta_{d+1} = dw \wedge (B\ d^d y) +\cdots$
on $Y$ is defined by $\int_Y \beta_{d+1} : = dw \int B d^dy$. However, we have $d_x ( dw \wedge B d^d y)=-dw \wedge (\partial B) d^{d+1} x$
because $d_x$ and $dw$ anti-commute. Because of this minus sign, we needed the change of sign in the Stokes theorem from $-2 \pi \ii  \int_{Y_w}  I^0_{d+1}$
to $+2 \pi \ii  \int_{X_w} d_x I^0_{d+1}$.

By using the fact that 
$(d_w+d_x)I^0_{d+1}=dI^0_{d+1}=I_{d+2}$, we finally get
\beq
\nabla  \log \CT (X_w)  = 2\pi \ii \int_{X_{w}} I_{d+2}. 
\eeq
This is the desired result.

We have used $I^0_{d+1}$ in the above computation. But one can also perform computation only by using $I_{d+2}$
by comparing the manifolds corresponding to $ \bra{X_{w+dw}} Y_{dw} \ket{X_w} $ and $ \bra{X_w} \et{X_w} $.

\section{Examples} \label{sec:example}
Here we give a few examples of topological phases of matter.

\subsection{Integer quantum Hall state}
Let us consider a $d+1=2+1$ dimensional fermion which is coupled to a $\U(1)$ (electromagnetic) gauge field $A$.
We take the spatial manifold $Y$ to be a torus. We denote this torus as $T^2_Y$.
The coordinates on $T^2_Y$ are denoted as $(y^1,y^2)$ with periodic conditions
\beq
(y^1, y^2) \simeq (y^1+1, y^2) \simeq (y^1, y^2+1).
\eeq
One can introduce an arbitrary metric on this torus, but 
we assume that the metric is flat for simplicity.

We consider two parameter family of $\U(1)$ gauge fields given by
\beq
A = - 2\pi \ii (  w_1 dy^1+ w_2 dy^2),
\eeq
where $w=(w_1,w_2) \in W$ are parameters, and our convention is such that we take gauge field and connections to be anti-hermitian.  
Two points $(w_1, w_2)$ and $(w_1+1, w_2)$ on the parameter space are
just related by the gauge transformation by $e^{2\pi \ii y^1}$, so we identify $(w_1, w_2) \simeq (w_1+1, w_2)$.
In the same way we identify $(w_1, w_2) \simeq (w_1, w_2+1)$. Therefore, the parameter space $W$ is also a torus $T^2$ which we denote as $T^2_W$
to distinguish it from the spatial torus $T^2_Y$. The total space of the fiber bundle is $F=T^2_W \times T^2_Y$.

Now let us apply the formula \eqref{eq:Bcurvature}.
The $F$ in that formula is given by
\beq
F=dA =(d_w+d_y)A=  - 2\pi \ii (  dw_1 \wedge dy^1+ dw_2 \wedge dy^2).
\eeq
The $\hat{A}(R)$ is just $1$ because the metric is assumed to be flat and constant. So we get
\beq
\CF=   -2\pi \ii \int_{Y_w}  \frac{1}{2} \left( \frac{\ii F}{2\pi} \right)^2=  -2\pi \ii  \int_{Y_w} - dw_1 \wedge dw_2 \wedge dy^1 \wedge dy^2=2\pi \ii  dw_1  \wedge dw_2.
\eeq
In particular, the first chern class of this curvature is given by
\beq
c_1(\CF)=\frac{\ii \CF }{2\pi} =-dw_1 \wedge dw_2.
\eeq
Integration of $c_1(\CF)$ over $T^2_W$ is equal to $-1$. This means that the ground state takes values in a nontrivial line bundle.
The above computation is essentially the same as that explained in \cite{Witten:2015aoa}.

The value of $\int_{T^2_W} c_1(\CF)$ is known to be the same as the integer appearing in the integer quantum Hall effect~\cite{Thouless:1982zz}.
Thus we recovered the fact that one massive fermion with mass parameter $m<0$ (or more precisely the difference of the theories with $m>0$ and $m<0$)
gives one unit of integer of the integer quantum Hall effect. If we have $k$ fermions with $m<0$, we get the integer as $\int_{T^2_W} c_1(\CF) =-k$.

\subsection{Majorana chain}
Although the title is ``Majorana"chain, we consider a Dirac fermion for simplicity because we have been treating Dirac fermions in this paper.
Let us consider a $d+1=1+1$ dimensional fermion which transforms under the Lorentz group as $\Pin^-(1,1)$ (or $\Pin^-(2)$ in Euclidean signature).
There are no gauge fields.
This is the model considered in \cite{Fidkowski:2009dba} (see also \cite{Witten:2015aba}).
One Dirac fermion is just two copies of Majorana fermions.

In this model, the situation is subtle, although computations are simple.
There is only one manifold with $d=1$ without boundary, namely a circle $Y=S^1$, and it is just parametrized by the circumference $L$ up to diffeomorphisms.
This $S^1$ is parametrized by a coordinate $y$ with $y \simeq y+1$ and the metric is $ds^2=L^2 dy^2$.

There is a diffeomorphism $\varphi : y \to -y$ which acts on the fermion as
\beq
\Psi(\tau,y) \to \bar{\gamma}\gamma^y \Psi(\tau,-y)=- \ii \gamma^\tau \Psi(\tau,-y),\label{eq:reflection}
\eeq
where $\bar{\gamma}=\ii^{-1} \gamma^\tau \gamma^y$.
Under the diffeomorphism $\varphi$, the metric is invariant. In general, if a metric $g$ is invariant under some diffeomorphism, $\varphi_* g=g$,
that metric may be considered to be at an ``orbifold singularity" of the moduli space of metrics.
(A famous example of such an orbifold singularity is given by the point $\tau=\ii$ in the moduli space of complex structures of a two dimensional torus $T^2$.)
In this sense of the orbifold singularity, any metric on the circle $S^1$ is always at the orbifold singularity of $\varphi$. 
This fact makes the discussion of this case a bit different from the one based on Berry phases.
This is the subtlety of this model. Nevertheless, the formula \eqref{eq:Beta} still makes a certain sense as we will see below.

Let us use the standard APS boundary condition for all the nonzero modes on $Y=S^1$. The standard APS boundary condition is completely 
invariant under diffeomorphisms, so we forget about those nonzero modes. If we consider the anti-periodic condition for $\Psi$, there are no zero modes
and we have the unique ground state $\ket{\Omega,U_Y}$. However, for the periodic boundary condition,
there are two zero modes $\Psi_{\pm,0}$, one with positive chirality and one negative chirality with respect to $\gamma^\tau$.
We consider a generalized APS boundary condition as
\beq
\Psi_{-,0}=T_0 \Psi_{+,0}
\eeq
where $T_0$ is a complex number with $|T_0|=1$.

However, under the diffeomorphism \eqref{eq:reflection}, one can see that the generalized APS boundary condition changes as
\beq
T_0 \to T'_0=-T_0
\eeq
and hence
\beq
\ket{\Omega,T_0}  \to \ket{\Omega,T'_0} = - \ket{\Omega,T_0}.
\eeq
Therefore, the state $\ket{\Omega,T_0} $ is not invariant under $\varphi$. This fact should be interpreted as a kind of nontriviality of $\CL$.

We can construct a manifold $Y_\gamma$ by starting from $[0,1] \times Y$ and gluing $\tau=0$ and $\tau=1$ using the diffeomorphism $\varphi$.
Then we get a Klein bottle $K$. We denote the Klein bottle with the anti-periodic and periodic boundary conditions in the direction of $Y=S^1$
as $K_A$ and $K_P$, respectively. Then, a straightforward computation (using the explicit mode expansion)
gives
\beq
 \exp({-2 \pi \ii \eta_{K_A}})=+1,~~~~~~ \exp({-2 \pi \ii \eta_{K_P}})=-1.
\eeq
These results are completely in accord with the above discussion about $\ket{\Omega,U_Y}$ and $\ket{\Omega,T_0}$.

If we take two copies of Dirac fermions (or four copies of Majorana), the above phase ambiguity vanishes and the line bundle $\CL$
is trivial in that sense. But that does not means that the theory with four Majorana fermions is trivial, as remarked at the end of Sec.~\ref{sec:detline}.
We need eight Majorana fermions to make the low energy theory trivial \cite{Fidkowski:2009dba,Witten:2015aba}.

In the same way, we can also consider topological superconductors in $d+1=3+1$ dimensions.
There, the bundle $\CL$ can be trivialized when there are $\nu$ Majorana fermions which is a multiple of 8~\cite{Hsieh:2015xaa,Witten:2015aba}. The low energy
theory becomes completely trivial  
if $\nu$ is a multiple of 16 
\cite{Fidkowski:2013jua,Wang:2014lca,KitaevCollapse,Metlitski:2014xqa,Morimoto:2015lua, Kapustin:2014dxa,Witten:2015aba, Tachikawa:2016xvs, Witten:2016cio}.
It would be interesting to compute very explicitly the structure of $\CL$ over $W$ by using techniques analogous to \cite{Hsieh:2015xaa}.

\section*{Acknowledgments}
The work of K.Y is supported by World Premier International Research Center Initiative
(WPI Initiative), MEXT, Japan. 

\appendix

\section{Implications for anomalies}\label{sec:appA}

Here we review implications of the Dai-Freed theorem for anomalies~\cite{Witten:2015aba,Witten:2016cio} to demonstrate the power of the theorem. 
The notations follow those of the Introduction.

Let us consider a $d$-dimensional manifold $Y$ which is either an even dimensional spin manifold or odd dimensional $\pin^\pm$ manifold.
On this space $Y$, we can consider a chiral fermion $\psi_{+}$ with positive chirality and the chiral Dirac operator is $ \CD^{-+}_Y$.
The partition function of the chiral fermion is given as
\beq
\det \CD^{-+}_Y.
\eeq
However, it is not straightforward to make sense of this expression.
The $\CD^{-+}_Y$ is a linear map from the space $H_+(Y)$ to $H_-(Y)$.
Because $H_+(Y)$ and $H_-(Y)$ are different vector spaces, the determinant of $\CD^{-+}_Y$ does not give a number in $\bC$, but
physically we need a partition function to take values in $\bC$. However, because these vector spaces have a natural hermitian inner product,
the absolute value
\beq
|\det \CD^{-+}_Y| =\sqrt{\CD^{+-}_Y\CD^{-+}_Y}
\eeq
is well-defined (after some appropriate regularization).

Then we might try to define the fermion partition function just by this absolute value $|\det \CD^{-+}_Y| $. However,
this is not a smooth function of the metric and gauge field. When we change the metric and gauge field, some of the eigenvalues of $\CD^{+-}_Y\CD^{-+}_Y$
hit zero. At that point, $|\det \CD^{-+}_Y| $ is not smooth. This is analogous to the fact that functions like
 $|w|=\sqrt{w^2}$ \cite{Witten:1982fp} or $\sqrt{w^2+w'^2}$ \cite{AlvarezGaume:1983cs} are not smooth functions of $w$ and $w'$, where $w$ and $w'$ correspond to parameters
 of metric and gauge field in our case.
When we compute correlation functions of the energy-momentum tensor and the current coupled to the gauge field
by using the partition function, we take functional derivatives of the partition function. So 
at least we need to require the condition that the partition function is a smooth function of the metric and gauge field.

Instead of considering just the absolute value, we consider the following quantity. We pick up a manifold $X$ whose boundary is $Y=\partial X$ (assuming this is possible).
Near the boundary, $X$ is assumed to have the form $(-\tau_0, 0] \times Y$.
Then we consider
\beq
|\det \CD^{-+}_Y| \exp(-2\pi \ii \eta(U_Y)).\label{eq:smth}
\eeq
Here $\eta(U_Y)$ is the eta-invariant computed by using the standard APS boundary condition specified by $U_Y$ in \eqref{eq:sAPS}.
Neither $|\det \CD^{-+}_Y| $ nor $\exp(-2\pi \ii \eta(U_Y))$ is smooth, but the above combination is smooth.
Let us see this. The standard APS boundary condition becomes ill-defined when some of the eigenvalues hit zero.
Near such a point, we pick up an arbitrary generalized APS boundary condition $T$ 
(or more precisely this is a smooth function of metric and gauge field, $T(g,A)$) which is locally well-defined near that point.
In other words, $T$ is chosen in such a way that it is independent of the behavior of small eigenvalues.
Now we use Theorem~1. By \eqref{eq:DFth1}, we can write
\beq
|\det \CD^{-+}_Y| \exp(-2\pi \ii \eta(U_Y)) &=|\det \CD^{-+}_Y|  \det(U_YT^{-1})\exp(-2\pi \ii \eta(T)) \nonumber \\
&=\det( \CD_Y^{-+}T^{-1})\exp(-2\pi \ii \eta(T))
\eeq
where we have used the definition of $U_Y$ given in \eqref{eq:sAPS}.
This is now manifestly smooth; the absolute value $|\det \CD^{-+}_Y|$ is cancelled, and both $\det( \CD_Y^{-+}T^{-1})$ and $\exp(-2\pi \ii \eta(T))$ are smooth.

From now the discussion depends on whether $Y$ is an even dimensional spin manifold or odd dimensional $\pin^\pm$ manifold. 
We first consider the case of an even dimensional spin manifold. Although \eqref{eq:smth} is smooth, it depends on the choice of $X$.
In particular, if we change the metric and gauge field on $X$ while fixing them near the boundary $Y$, the value of $\eta$ changes.
We want to cancel this dependence.
This can be done by considering the Chern-Simons action of the background fields. 
We define $I^0_{d+1}$ by
\beq
dI^0_{d+1} &=I_{d+2}= \left. \hat{A}(R) \tr \exp( \frac{\ii F}{2\pi}) \right|_{d+2},
\eeq
where $ \hat{A}(R)$ is the $\hat{A}$ genus of the Riemann tensor $R$ and $F$ is the curvature of the background gauge field.
Then the combination
\beq
\exp(-2\pi \ii \eta(U_Y)) \exp(-2 \pi \ii \int_{X} I^0_{d+1} ) \label{eq:invcom}
\eeq
is invariant under small continuous change of the metric and background gauge field on $X$ if they are held fixed near the boundary.
This is a consequence of Theorem~3. If the metric and gauge field near the boundary $\partial X=Y$ are held fixed, the connection $\nabla$ of
Theorem~3 is just the ordinary exterior derivative on the parameter space of metric and gauge field, 
and \eqref{eq:DFth3} implies that \eqref{eq:invcom} is independent of the metric and gauge field inside $X$.
Therefore the combination
\beq
Z_\psi=|\det \CD^{-+}_Y| \exp(-2\pi \ii \eta(U_Y)) \exp(-2 \pi \ii \int_X I^0_{d+1} ) \label{eq:chiralp}
\eeq
is independent of the metric and gauge field inside the $X$ if the boundary is held fixed, at least under small deformation.
Therefore, it only depends on the metric and gauge field on $Y$, at least under small deformation, and hence 
it has desired properties to be the partition function of the chiral fermion on $Y$.

However, \eqref{eq:chiralp} is gauge dependent. Under a gauge transformation $\delta_{\rm gauge}$, 
the standard descendant equations of the anomaly polynomial (see e.g., \cite{Weinberg:1996kr})
tell us that $I^0_{d+1}$ behaves as
\beq
\delta_{\rm gauge} I^0_{d+1} = d I^{1}_d.
\eeq
Then we get
\beq
\delta_{\rm gauge} \log Z_\psi = -2 \pi \ii \int_X dI^1_{d} = -2 \pi \ii \int_Y I^1_{d}. \label{eq:gvariation}
\eeq
This exactly reproduces the chiral anomaly we expect from one-loop perturbative computation~\cite{AlvarezGaume:1983ig}.\footnote{
From the path-integral point of view, the minus sign in \eqref{eq:gvariation} comes from the fact that path integral measure of a grassmann variable $\xi$
is such that $ d (e^{ \ii \alpha} \xi) =  (d \xi) e^{ -\ii \alpha}$.
}

Next, let us consider the case that the anomaly polynomial $I_{d+2}$ is zero. This includes the case of an odd dimensional $\pin^\pm$ manifold $Y$,
as well as the case of an even dimensional spin manifold $Y$ if the matter content is such that the perturbative anomaly is cancelled. 
In this case, we can use the definition of the partition function $Z_\psi$ of \eqref{eq:chiralp} with $I^0_{d+1}=0$, and there is no anomaly under small gauge transformation.

Even though \eqref{eq:chiralp} is invariant under small change of the metric and gauge field inside $X$, it may still have dependence on $X$ under a large change of $X$.
This can be studied by Theorem~2. We pick up another manifold $X'$ with $\partial X' = Y$, and then we can define $Z_\psi$ by using $X'$.
The difference between the two definitions is given by
\beq
 \exp(-2\pi \ii \eta_X(U_Y))  \exp(+2\pi \ii \eta_{X'}(U_Y)) = \exp(-2\pi \ii \eta_X(U_Y))  \exp(-2\pi \ii \eta_{ -X'}(U_{-Y})) \label{eq:dffp}
\eeq
where we used $\eta_{X'}(U_Y)=-\eta_{ -X'}(U_{-Y})$.
We can glue the manifold $X$ and $-X'$ together to get a manifold $X_{\rm com}$ without boundary. Then by using \eqref{eq:DFth2-1},
the phase difference \eqref{eq:dffp} is now given by
\beq
 \exp(-2\pi \ii \eta_{X_{\rm com}})  .
\eeq
This was proposed as the general formula for the global anomaly~\cite{Witten:2015aba,Witten:2016cio}.

In the current framework, a more traditional global anomaly \cite{Witten:1982fp, Witten:1985xe} is understood as follows.
We denote by $Y[g,A]$ the manifold $Y$ equipped with the metric $g$ and gauge field $A$.
Once we define $Z_\psi$ by \eqref{eq:chiralp} (with $I^0_{d+1}=0$), we can consider a change of the phase factor 
from $Y[g,A]$ to $Y[g',A']$ by extending $X$ by attaching a manifold $\Delta Y=[0,1] \times Y$.
The metric and gauge field of $\Delta Y$ at $\{0\} \times Y$ and $\{1\} \times Y$ are given by $[g,A]$ and $[g',A']$, respectively.
The $X$ is glued to $\Delta Y$ at $\{0\} \times Y$. Now, the global anomaly in the sense of \cite{Witten:1982fp,Witten:1985xe}
is the change of the phase factor when $[g',A']$ is a transformation of $[g,A]$ by a combined diffeomorphism and gauge transformation $\varphi$ which we denote as
$[g',A'] = \varphi[g,A]$.
By repeated use of the general gluing theorem \eqref{eq:DFth2}, one can check that the change of the phase factor is given by 
$ \exp(-2\pi \ii \eta_{(S^1 \times Y)_\varphi})$, where $(S^1 \times Y)_\varphi$ is a manifold which is obtained by gluing $\{0\} \times Y$ and $\{1\} \times Y$ of 
$\Delta Y=[0,1] \times Y$
by using the twist by $\varphi$. This $ \exp(-2\pi \ii \eta_{(S^1 \times Y)_\varphi})$ reproduces the formula of \cite{Witten:1985xe}.

Finally, we consider a variation of $Z_\psi$ under small changes of the metric and gauge field on $Y$.
For simplicity we assume that perturbative anomaly is cancelled so that $I_{d+2}=0$. 
The fact that $ \exp(-2\pi \ii \eta(T))$ is invariant under small deformation inside $X$ suggests
that this quantity actually depends only on the boundary condition $T$, at least under small deformation. 
This statement essentially follows from \eqref{eq:DFth3}, because when the perturbative anomaly $I_{d+2}$ is zero, the connection $\nabla$ is flat
and $\CT=\exp(-2\pi \ii \eta_X(T))/ \det(T)$ is constant.
Then, when the Dirac operator is changed from $\CD_Y$ to $\CD_Y' =\CD_Y+\delta \CD_Y$,
we get
\beq
\exp(-2\pi \ii \eta(U_Y')) = \det(U_Y' U_Y^{-1})\exp(-2\pi \ii \eta(U_Y)) \label{eq:var}
\eeq
where $U_Y'$ is defined in terms of $\CD_Y' =\CD_Y+\delta \CD_Y$.
For the equation \eqref{eq:DFth1} to be valid, it was necessary that the boundary condition satisfies the condition \eqref{eq:gAPS}.
However, \eqref{eq:DFth1} makes sense as long as $T $ approaches to $U_Y$ sufficiently rapidly for high frequency modes.
In the present case where $T=U'_Y$, it may be realized by e.g., a heat kernel regularization 
$\delta \CD_Y \to e^{-\CD_Y^2/2\Lambda^2} \delta \CD_Y e^{-\CD_Y^2/2\Lambda^2} $.
Assuming this is done, \eqref{eq:var} gives\footnote{Our argument here is sketchy. More precise treatment might require the details of the Bismut-Freed connection; 
see \cite{Freed:1986hv} for the details of that connection. }
\beq
\delta \log Z_\psi = \tr \left( \frac{1}{\CD_Y^{-+}}  \delta \CD_Y^{-+} \cdot e^{- \CD_Y^{+-}\CD_Y^{-+}/\Lambda^2}  \right) \label{eq:pertcom}
\eeq
where for simplicity we used heat regularization also for the part $|\det \CD^{-+}_Y| $ as 
$\log |\det \CD^{-+}_Y| =\frac{1}{2}\int^\infty_{1/\Lambda^2} \frac{dt}{t} e^{-t \CD^{+-} \CD^{-+} }$.
The \eqref{eq:pertcom} is exactly what we expect in perturbative computation (in heat regularization).
This equation guarantees that $Z_\psi$ reproduces the same correlation functions as the ones computed in the standard perturbation theory.
In this sense, the definition \eqref{eq:chiralp} is physically sensible.

\section{Conventions for gamma matrices and chirality}\label{sec:convention}
Our conventions of the chirality of fermions in the orientable spin case with $d = 2n$ are as follows.
First of all, we want to use the relation of the Dirac operator between $X$ and $Y = \partial X$ as
 $\CD_X= \ii \gamma^\mu D_\mu =\ii \gamma^\tau ( \partial_\tau+\CD_Y)$, which is convenient for the calculations in the main text.
 The orientation on $X$ is defined by $dx^0 \wedge dx^1 \wedge \cdots \wedge dx^{2n}=d\tau \wedge dy^1 \wedge \cdots \wedge dy^{2n}$.
 We also want $\CD_Y$ to be given as $\ii \tilde{\gamma}^i D_i$.
 Here we denoted the gamma matrices in $Y$ as $\tilde{\gamma}^i~(i=1,2,\cdots, 2n)$. Then we get 
 \beq
 \tilde{\gamma}^i = - \ii \gamma^\tau \gamma^i.
 \eeq
 Furthermore, we want to use $\gamma^\tau$ as the chirality operator $\bar{\tilde{\gamma}}$ on the fermions on $Y$. We require
\beq
 \bar{\tilde{\gamma}} = \ii^{-n} \cdot \tilde{ \gamma}^1 \tilde{ \gamma}^2 \cdots \tilde{ \gamma}^{2n}.
\eeq
By this convention of chirality, 
the Atiyah-Singer index theorem says that the number of positive chirality zero modes of $\CD_Y$ minus the number of negative chirality zero modes is given by 
$\int I_{2n}$. (One can check this statement in the case that $Y$ is a product of Riemann surfaces as $Y = \Sigma \times \cdots \times \Sigma $ and 
then using the Riemann-Roch theorem.)
Then, by requiring $\gamma^\tau = \bar{\tilde{\gamma}}$ and denoting $\gamma^\tau=\gamma^0$, we get 
(in Euclidean signature with the orientation $dx^0 \wedge dx^1 \wedge \cdots \wedge dx^{2n}$ )
\beq
\gamma^0 \gamma^1 \cdots \gamma^{2n} = \ii^n. \label{eq:oddchirality}
\eeq
This is our convention for the representation of the Clifford algebra acting on the fermion $\Psi$ on $X$.

Unfortunately, under the above conventions, 
the relation between the eta-invariant $\eta_X$ and the Chern-Simons invariant needs an unconventional minus sign factor
as
\beq
\eta_X = - \int_X I_{2n+1} + {\rm const.}  \label{eq:etaCS}
\eeq
where $X$ is assumed to have no boundary $\partial X = \varnothing$ in this equation.
To see this, let us actually derive it. 

Consider a $2n+2$-dimensional manifold $\dualX = \bR \times X$ and a chiral fermion $\dualPsi$ on this space. 
The gamma matrices in this space are denoted as $\Gamma^M$,
and the coordinate for the time direction $\bR$ is denoted as $\dualt$. 
We use Lorentzian signature for $\dualt$, and the corresponding gamma matrix $\Gamma^\dualt$ satisfies $(\Gamma^\dualt)^2=-1$.
The chirality on $\dualX$ is defined by $\bar{\Gamma} = \ii^{-(n+1)} ( \ii \Gamma^\dualt) \Gamma^0 \Gamma^1 \cdots \Gamma^{2n}$.
The Lagrangian is
\beq
\CL=-\bar{\dualPsi}  \Gamma^M D_M  \left( \frac{1+\bar{\Gamma} }{2} \right) \dualPsi
\eeq
where $\bar{\dualPsi} = \dualPsi^\dagger \ii \Gamma^\dualt$. 

In this theory, we can consider the axial current
\beq
J^M = \ii \bar{\dualPsi} \Gamma^M \left( \frac{1+\bar{\Gamma} }{2} \right) \dualPsi.
\eeq
By the standard way, one can check that this axial current has an anomaly given by
\beq
(\nabla_M J^M) \cdot d \dualt \wedge dx^0 \wedge  \cdots \wedge dx^{2n} = I_{2n+2},
\eeq
Then the charge $Q = \int_X J^\dualt$ behaves as
\beq
Q(\dualt_1) - Q(\dualt_0) = \int_{[\dualt_0, \dualt_1] \times X}  I_{2n+2}.\label{eq:anomalouschange}
\eeq
On the other hand, the charge $Q$ is given by the eta-invariant. To see that, notice that the Hamiltonian is given by
\beq
H_X=\dualPsi^\dagger \ii \Gamma^\dualt \Gamma^\mu D_\mu  \left( \frac{1+\bar{\Gamma} }{2} \right) \dualPsi 
=\dualPsi^\dagger \CD_X \dualPsi \label{eq:ham}
\eeq
where the relation of gamma matrices is defined by
\beq
\gamma^\mu = \Gamma^\dualt \Gamma^\mu  \left( \frac{1+\bar{\Gamma} }{2} \right).
\eeq
One can check that 
\beq
\gamma^0 \gamma^1 \cdots \gamma^{2n}=\ii^{n}  \left( \frac{1+\bar{\Gamma} }{2} \right).
\eeq
This is precisely \eqref{eq:oddchirality} when
projected to the positive chirality space $(1+\bar{\Gamma})/2=1$. On the other hand, the charge $Q$ is
\beq
Q =\int_X \dualPsi^\dagger  \dualPsi = \frac{1}{2}\int_X ( \dualPsi^\dagger  \dualPsi - \dualPsi \dualPsi^\dagger).
\eeq
Therefore, in the vacuum state $\ket{{\rm VAC}} $ of the Hamiltonian \eqref{eq:ham}, we get
\beq
Q \ket{{\rm VAC}} = - \frac{1}{2} \sum_{\lambda_X} \sign(\lambda) \ket{{\rm VAC}}= - \eta_X \ket{{\rm VAC}} ,\label{eq:axialcharge}
\eeq
where we assumed that $\CD_X$ has no zero modes for simplicity.
Then, \eqref{eq:anomalouschange} and \eqref{eq:axialcharge} implies \eqref{eq:etaCS}.

We remark that $Q$ behaves smoothly under the change of metric and gauge field, but the vacuum $\ket{{\rm VAC}} $ changes discontinuously
when some of $\lambda_X$ hit zero, and that gives the discontinuous change of the eta-invariant by integer values. 

\bibliographystyle{JHEP}
\bibliography{ref}

\end{document}